\documentclass{emulateapj}
\usepackage{color}
\begin{document}

\title{Early Science with the Large Millimeter Telescope: Exploring the effect of AGN activity on the relationships between molecular gas, dust, and star formation}
\author{Allison Kirkpatrick\altaffilmark{1}, Alexandra Pope\altaffilmark{1}, Itziar Aretxaga\altaffilmark{2}, Lee Armus\altaffilmark{3}, Daniela Calzetti\altaffilmark{1}, George Helou\altaffilmark{4}, Alfredo Monta\~{n}a\altaffilmark{2}, Gopal Narayanan\altaffilmark{1}, F. Peter Schloerb\altaffilmark{1},
Yong Shi\altaffilmark{5}, Olga Vega\altaffilmark{2}, Min Yun\altaffilmark{1}}
\altaffiltext{1}{Department of Astronomy, University of Massachusetts, Amherst, MA 01002, USA, kirkpatr@astro.umass.edu}
\altaffiltext{2}{Instituto Nacional de Astrof\'{i}sica, Optica y Electr\'{o}nica, Apdos. Postales 51 y 216, C.P. 72000 Puebla, Pue., Mexico}
\altaffiltext{3}{Spitzer Science Center, California Institute of Technology, MS 220-6, Pasadena, CA 91125, USA}
\altaffiltext{4}{Infrared Processing and Analysis Center, California Institute of Technology, Pasadena, CA 91125, USA}
\altaffiltext{5}{School of Astronomy and Space Science, Nanjing University, Nanjing, 210093, China}

\begin{abstract}
The molecular gas, H$_2$, that fuels star formation in galaxies is difficult to observe directly. As such, the ratio of $L_{\rm IR}$ to $L_{\rm CO}^\prime$
is an observational estimation of the star formation rate compared with the amount of molecular gas available to form stars, which is related to the star
formation efficiency and the inverse of the gas consumption timescale.
We test what effect an IR luminous AGN has on the ratio $L_{\rm IR}/L_{\rm CO}^\prime$ in a sample of 24 intermediate redshift galaxies from the 5 mJy Unbiased {\it Spitzer} Extragalactic Survey (5MUSES). We obtain new CO(1-0) observations with the Redshift Search Receiver on the Large Millimeter Telescope. 
We diagnose the presence and strength of an AGN using {\it Spitzer} IRS spectroscopy. We find that removing the AGN contribution to $L_{\rm IR}^{\rm tot}$ results in a mean $L_{\rm IR}^{\rm SF}/L_{\rm CO}^\prime$ for our entire sample consistent with the mean $L_{\rm IR}/L_{\rm CO}^\prime$ derived for a large sample of star forming galaxies from $z\sim0-3$. We also include in our comparison the relative amount of polycyclic aromatic hydrocarbon emission for our sample and a literature sample of local and high redshift Ultra Luminous Infrared Galaxies and find a consistent trend between $L_{6.2}/L_{\rm IR}^{\rm SF}$ and
$L_{\rm IR}^{\rm SF}/L_{\rm CO}^\prime$, such that small dust grain emission decreases with increasing $L_{\rm IR}^{\rm SF}/L_{\rm CO}^\prime$ for both local and high redshift dusty galaxies.
\end{abstract}

\section{Introduction}
Star formation is one of the main internal driving forces of galaxy evolution, resulting in the chemical enrichment of a galaxy, the heating of the interstellar medium (ISM), and indirectly, the production of dust through the winds of dying stars.
Star formation converts a galaxy's molecular gas into stars through multiple complicated processes including gas accretion and the collapse and cooling of molecular clouds. Although the star formation process itself is intricate, the overall conversion of gas into stars can be expressed simply by the Schmidt-Kennicutt (SK) law which directly relates the molecular gas content to the star formation rate (SFR) through a power-law equation, $\Sigma_{\rm SFR} \propto \Sigma_{\rm gas}^{\alpha}$, albeit with significant scatter \citep{schmidt1959,kennicutt1998,kennicutt2012}.

The bulk of the present day stellar mass was formed at a peak epoch of star formation, from $z\sim1-3$ \citep[and references therein]{madau2014}.
During this era, the buildup of stellar mass was dominated by dusty galaxies referred to as Luminous Infrared Galaxies (LIRGs, $L_{\rm IR}=10^{11}-10^{12}L_\odot$) and Ultra Luminous Infrared Galaxies (ULIRGs, $L_{\rm IR}>10^{12}L_\odot$) \citep[e.g.,][]{murphy2011a}. In the past two decades, the spate of far-IR/submillimeter space-based and ground-based telescopes have enabled astronomers to simultaneously study the star formation, through infrared (IR) emission, and molecular gas, through CO emission, of dusty galaxies out
to redshifts of $z\sim4$ \citep[and references therein]{carilli2013}. The IR luminosity, $L_{\rm IR}$, is an ideal measure of the SFR for dusty galaxies as it is the integrated emission from the dust, presumably heated by star formation. On galaxy-wide scales, CO traces the molecular hydrogen which is difficult to observe directly; a conversion factor, $\alpha_{\rm CO}$, is used to relate the CO luminosity directly to the H$_2$ mass \citep[and references therein]{bolatto2013}.

In the past few years, a ``galaxy main sequence'' has been empirically determined for local and high redshift galaxies; a tight relationship holds between SFR and stellar mass, and this relationship evolves with redshift  \citep[e.g.,][]{noeske2007,elbaz2011}. Galaxies that lie above the main sequence, that is, galaxies that have an enhanced SFR for a given stellar mass, are designated ``starbursts'' in this parameter space, as they are thought to be undergoing a short-lived burst of star formation, likely triggered by a major merger. The rate at which a galaxy can form stars is limited by the amount of molecular gas present.
\citet{shi2011} proposed an extended SK law which relates the specific star formation rate ($\Sigma_{\rm SFR}/\Sigma_{\rm gas}$) to the stellar mass surface density, suggesting that the existing stellar population may play a role in regulating the amount of star formation. The authors apply their extended SK law to an analytical model of gas accretion and find that it accurately reproduces the galaxy main sequence.

A dichotomy between starbursts and normal star forming galaxies may also be observed by comparing $L_{\rm IR}$ with $L_{\rm CO}^\prime$. First, there may be a ``normal'' rate of star formation measured in undisturbed disk galaxies for a given amount of molecular gas. Then, there is an enhanced starburst mode, where a galaxy has a higher $L_{\rm IR}$ than expected for a given $L_{\rm CO}^\prime$, possibly triggered by a major merger funneling gas towards the inner regions of a galaxy \citep[e.g.,][]{downes1998,solomon2005,daddi2010,genzel2010}. This dichotomy has a direct effect on the calculation of the H$_2$ mass, since $\alpha_{\rm CO}$ is proposed to be a factor of $\sim5$ lower for starbursting galaxies, due to significant amounts of CO residing in the inter-cloud medium. Correctly identifying starbursts is critical for calculating accurate gas masses.

It now appears that every massive galaxy hosts a supermassive black hole at the
center, implying that all galaxies have gone through an active galactic nuclei (AGN) phase, and some LIRGs and ULIRGs show signs of concurrent black hole growth and star formation \citep[e.g.,][]{sajina2007,pope2008,coppin2010,santos2010,wu2010,petric2011,kirkpatrick2012}. In the classical picture of galaxy evolution, a starburst is triggered by
a major merger, and this phase can be followed by an AGN phase, implying that an obscured growing AGN may be observable during a galaxy's starburst phase \citep[e.g.,][]{sanders1996}.
If present, an AGN can dramatically influence the internal evolution of a galaxy by heating the dust, expelling the gas, and ultimately quenching the star formation through 
feedback mechanisms \citep[e.g.,][]{alexander2012}. 
The effect of an obscured AGN on the ISM can be probed through IR observations. In the mid-IR, star forming galaxies have prominent PAH features arising from photodissociation regions (PDRs), but AGN emission can dilute these features, leaving a warm dust power-law
continuum \citep{weedman2004,siebenmorgen2004}. Radiation from the AGN can heat the dust in the ISM to temperatures $\gtrsim100\,$K, producing a
significant contribution ($\gtrsim50\%$) to the far-IR emission \citep[e.g.,][]{mullaney2011,kirkpatrick2012}.

The degree of scatter in the relationship between $L_{\rm IR}$ and $L_{\rm CO}^\prime$, often attributed to the two different modes of star formation,
is possibly affected by the contribution from an AGN to $L_{\rm IR}$. 
\citet{evans2001} observe high $L_{\rm IR}/L_{\rm CO}^\prime$ in QSOs and infer that these ratios might be boosted by an AGN contribution to $L_{\rm IR}$. \citet{evans2006} build on this study by using HCN as a tracer of star formation in QSOs. Using IRAS bright galaxies, the authors determine the median HCN/CO ratio for normal star forming galaxies and use this, combined with $L_{\rm IR}/L_{\rm CO}^\prime$ and $L_{\rm IR}/L_{\rm HCN}^\prime$, to statistically correct $L_{\rm IR}$ of QSOs for an AGN contribution. An alternate approach is to use $L_{\rm FIR}$ (40-500\,$\mu$m or 42.5-11.5\,$\mu$m) instead of $L_{\rm IR}$ (8-1000\,$\mu$m), as dust emission in this regime should come primarily from heating by young stars. \citet{xia2012} conclude that $L_{\rm FIR}/L_{\rm CO}^\prime$ in a sample of 17 QSOs is similar to the ratios in local ULIRGs (many of which are known to host obscured AGN), and \citet{evans2006} finds high $L_{\rm FIR}/L_{\rm CO}^\prime$ ratios in QSOs relative to the IRAS galaxies.

In this paper, we build upon observations in the high redshift and local Universe with a study of intermediate redshift galaxies ($z=0.04-0.35$) from the 5 mJy Unbiased {\it Spitzer} Extragalactic Survey (5MUSES; P.I. George Helou). These galaxies have extensive IR data from the {\it Spitzer Space Telescope} and {\it Herschel Space Observatory}, allowing us to accurately measure $L_{\rm IR}$ and quantify the contribution from an AGN.
We complement the existing IR data with new CO(1-0) observations from the Redshift Search Receiver, which has a large bandwidth of 38\,GHz and a resolution of 100\,km/s, on the Large Millimeter Telescope Alfonso Serrano, enabling us to explore how much of the scatter in the $L_{\rm IR}-L_{\rm CO}^\prime$ relation is due to an AGN contribution to $L_{\rm IR}$. We discuss the details of our sample and observations in Section 2, the effect of an AGN on the relationships between $L_{\rm IR}$, PAHs, and CO(1-0) emission in Section 3, and summarize our findings in Section 4.

Throughout this paper, we adopt a flat cosmology with $H_0=71\,$km\,s$^{-1}$\,Mpc$^{-1}$, $\Omega_m=0.27$, and $\Omega_\Lambda=0.73$.

\section{Data}
\begin{deluxetable*}{lccc cccc}
\tablecolumns{8}
\tablecaption{CO(1-0) Measurements of the 5MUSES sample\label{properties}}
\tablehead{\colhead{ID} & \colhead{RA} & \colhead{Dec} & \colhead{$z$\tablenotemark{a}} & \colhead{rms\tablenotemark{b}} & \colhead{FWHM} & \colhead{$S_{\rm CO}\Delta v$\tablenotemark{c}} & \colhead{$L_{\rm CO}^\prime$\tablenotemark{c}}\\
\colhead{} & \colhead{(J2000)} & \colhead{(J2000)} & \colhead{(CO)} & \colhead{(mK)} & \colhead{(km s$^{-1}$)} & \colhead{(Jy km s$^{-1}$)} & \colhead{($10^9$\,K\,km/s\,pc$^2$)}}
\startdata
5MUSES-200	& 16:12:50.9		& +53:23:05.0	& 0.043	& 0.70	& \nodata 	& $<2.53$		 	& $<0.21$ \\
5MUSES-179	& 16:08:03.7		& +54:53:02.0	& 0.053	& 0.32	& 348 $\pm$ 20	& 11.3 $\pm$ 0.93	& 1.46 $\pm$ 0.12 \\
5MUSES-169	& 16:04:08.3		& +54:58:13.1	& 0.064	& 0.72	& 306 $\pm$ 67	& 4.47 $\pm$ 1.08	& 0.84 $\pm$ 0.20  \\
5MUSES-105	& 10:44:32.9		& +56:40:41.6	& 0.068	& 0.84	& 251 $\pm$ 20	& 7.19 $\pm$ 1.27	& 1.54 $\pm$ 0.27 \\
5MUSES-171	& 16:04:40.6		& +55:34:09.3	& 0.078	& 0.87	& 345 $\pm$ 48	& 7.60 $\pm$ 1.45	& 2.14 $\pm$ 0.41 \\
5MUSES-229	& 16:18:19.3		& +54:18:59.1	& 0.082	& 0.72	& 386 $\pm$ 33	& 11.3 $\pm$ 1.48	& 3.53 $\pm$ 0.46 \\
5MUSES-230	& 16:18:23.1		& +55:27:21.4	& 0.084	& 0.87	& 305 $\pm$ 50	& 6.38 $\pm$ 1.25	& 2.09 $\pm$ 0.41  \\
5MUSES-234	& 16:19:29.6		& +54:18:41.9	& 0.100	& 0.70	& 217 $\pm$ 25	& 6.24 $\pm$ 0.69	& 2.93 $\pm$ 0.33 \\
5MUSES-132	& 10:52:06.6		& +58:09:47.1	& 0.117	& 1.30	& 181 $\pm$ 44	& 4.28 $\pm$ 0.92	& 2.77 $\pm$ 0.60 \\
5MUSES-227	& 16:17:59.2		& +54:15:01.3	& 0.134	& 0.24	& 247 $\pm$ 26	& 1.17 $\pm$ 0.34	& 1.00 $\pm$ 0.29 \\
5MUSES-141	& 10:57:05.4		& +58:04:37.4	& 0.140	& 0.48	& 203 $\pm$ 29	& 3.06 $\pm$ 0.62	& 2.86 $\pm$ 0.58 \\
5MUSES-158	& 16:00:38.8		& +55:10:18.7	& 0.145	& 0.73	& 458 $\pm$ 78	& 5.17 $\pm$ 1.23	& 5.21 $\pm$ 1.24 \\
5MUSES-225	& 16:17:48.1		& +55:18:31.1	& 0.145	& 0.46	& \, 400 $\pm$ 146	& 4.52 $\pm$ 0.57	& 4.55 $\pm$ 0.57  \\
5MUSES-273	& 16:37:31.4		& +40:51:55.6	& 0.189	& 0.34	& 241 $\pm$ 69	& 1.81 $\pm$ 0.40	& 3.16 $\pm$ 0.69 \\
5MUSES-136	& 10:54:21.7		& +58:23:44.7	& 0.204	& 0.36	& 477 $\pm$ 78	& 2.30 $\pm$ 0.55	& 4.69 $\pm$ 1.11 \\
5MUSES-294	& 17:12:32.4		& +59:21:26.2	& 0.210	& 0.38	& \nodata	& $<1.60$			& $<3.47$ \\
5MUSES-216	& 16:15:51.5		& +54:15:36.0	& 0.215	& 0.27	& 331 $\pm$ 60	& 1.82 $\pm$ 0.36	& 4.14 $\pm$ 0.83 \\
5MUSES-194	& 16:11:19.4		& +55:33:55.4	& 0.224	& 0.54	& \nodata		& $<2.62$	& $<6.49$ \\
5MUSES-249	& 16:22:14.8		& +55:06:14.2	& 0.237	& 0.41	& \nodata	& $<1.72$			& $<4.80$ \\
5MUSES-250 	& 16:23:13.1		& +55:11:11.6	& 0.237	& 0.48	& 599 $\pm$ 40	& 5.56 $\pm$ 0.68	& 15.5 $\pm$ 1.89\\
5MUSES-275	& 16:37:51.4		& +41:30:27.3	& 0.286	& 0.61	& \, 950 $\pm$ 163	& 7.77 $\pm$ 1.38	& 32.0 $\pm$ 5.68\\
5MUSES-313	& 17:18:52.7		& +59:14:32.1	& 0.322	& 0.47	& \nodata	& $<2.09$ 			& $<10.1$ \\
5MUSES-156	& 15:58:33.3		& +54:59:37.2	& 0.340	& 0.47	& \nodata	& $<2.17$			& $<12.8$ \\
5MUSES-101	& 10:41:59.8		& +58:58:56.4	& 0.360	& 0.32	& \nodata  	& $<1.34$			& \,\,\,\,\,\, \,\,\,\,\,\,\,\,\,\,\,$<8.91$ 
\tablenotetext{}{Sources are ordered by increasing redshift. }
\tablenotetext{a}{Redshift determined by fitting center of CO(1-0) line. Redshift errors scale with S/N, with the typical error of 0.0003 corresponding to S/N$\sim$5. For the sources with 3$\sigma$ upper limits, we list $z_{\rm IR}$.}
\tablenotetext{b}{The rms is determined over the entire 38\,GHz spectrum.}
\tablenotetext{c}{3$\sigma$ upper limits are listed for six sources where we were not able to detect a line.}
\end{deluxetable*}

\subsection{5MUSES Sample}
5MUSES is a {\it Spitzer} IRS mid-IR spectroscopic survey of 330 galaxies selected from the SWIRE and {\it Spitzer} Extragalactic First Look Survey fields \citep[details in][]{wu2010}. It is a flux limited sample selected at 24\,$\mu$m using MIPS observations from the {\it Spitzer Space Telescope}, with $S_{24}>5\,$mJy. Crucially, 5MUSES is a representative sample at intermediate redshift (the median redshift of the sample is 0.14, 1.8 Gyr ago) of galaxies with $L_{\rm IR}\sim10^{10}-10^{12}\,L_\odot$, bridging the gap between large samples of local star forming galaxies, LIRGs, and ULIRGs and high redshift observations of LIRGs and ULIRGs.
Complete details of the  {\it Spitzer} data reduction are found in \citet{wu2010}. For the present study, we make use of the {\it Spitzer} IRS short-low (SL) spectroscopy, spanning the range $5.5-14.5\,\mu$m.

A 24\,$\mu$m selection criteria could potentially bias the sample towards warmer sources. \citet{magdis2013} reports the far-IR observations of the 5MUSES sources with the {\it Herschel Space Observatory}. Of 188 5MUSES sources covered by the observations, 154 (82\%) are detected at 250\,$\mu$m. The authors also determine that the 250\,$\mu$m flux densities of the whole population of galaxies in the observed field with $S_{24}>5\,$mJy (the 5MUSES detection limit) and that of the 5MUSES sample is drawn from the same distribution, indicating that the dust temperatures of 5MUSES are not significantly warmer than other IR luminous galaxies. In Section \ref{sec:PAH}, we compare the 5MUSES sources to high redshift ULIRGs, also selected at 24\,$\mu$m, from the GOODS-{\it Herschel} survey. For the high redshift ULIRGs, \citet{pope2013} find no difference between 70\,$\mu$m selected sources ($\sim24\,\mu$m rest frame) and submillimeter selected sources in terms of PAH and CO luminosities. The high redshift ULIRGs in this work are drawn from a larger parent population of {\it Spitzer} IRS sources \citep{kirkpatrick2012}. \citet{kirkpatrick2012} find that 97\% of GOODS-{\it Herschel} sources with a 100\,$\mu$m detection also have a 24\,$\mu$m detection, ensuring that for high redshift LIRGs and ULIRGs, a 24\,$\mu$m selection criteria does not bias the selection towards warmer sources.

From the 5MUSES sample, we selected 24 sources that span a range of PAH equivalent widths \citep[EW; calculated for the full sample in][]{wu2010}, as PAH EW is
correlated with the mid-IR strength of an AGN \citep{armus2007}. All 24\,$\mu$m sources have spectroscopic redshifts determined from fitting the mid-IR spectral features. These sources were chosen for a pilot CO(1-0) study specifically because they exhibit a range of PAH strengths and because the spectroscopic redshifts are such that we can observe the CO(1-0) transition with the Large Millimeter Telescope (LMT). These sources were also selected according to $L_{\rm IR}$, such that every source is observable in a reasonable ($\lesssim90$ min) amount of time during the LMT Early Science phase. Our sample spans a redshift range of $z=0.04-0.36$ and $L_{\rm IR}=1.8\times10^{10}-1.3\times10^{12}\,L_\odot$. 
Details are listed in Table \ref{properties}.

\subsection{New LMT Observations}
The LMT is a 50\,m millimeter-wave radio telescope on Volc\'{a}n Sierra Negra, Mexico, at an altitude of 4600\,m \citep{hughes2010}. 
The high elevation allows for a median opacity of $\tau=0.1$ at 225\,GHz in the winter months. For the Early Science phase, the inner 32.5\,m of the 
primary reflector is fully operational. The primary reflector has an active surface and a sensitivity of 7.0\,Jy/K at 3\,mm. The pointing accuracy (rms) is $3^{\prime\prime}$ over the whole sky and 
1-2$^{\prime\prime}$ for small offsets ($<10^\circ$) from known sources.

During March and April 2014, we observed our 24 sources with the Redshift Search Receiver \citep[RSR;][]{erickson2007,chung2009}. The RSR is comprised of a dual beam, dual polarization system that simultaneously covers a wide frequency range of 73-111\,GHz in a single tuning with a spectral resolution of 100\,km/s.
This large bandwidth combined with a high spectral resolution is ideal for measuring the CO integrated line luminosities in high redshift sources. In our intermediate redshift sample, the large bandwidth allows us to observe the CO(1-0) transition for sources spanning a wide range of redshifts.
The beam size of the RSR is frequency dependent, such that $\theta_b =1.155 \lambda/32.5\,{\rm m}$. 
The beam size depends on the frequency of the observed CO(1-0) line, and at the median redshift, $z=0.145$, the RSR has a beam size of $22^{\prime\prime}$.

Typical system temperatures ranged from $87-106$\,K during our observations. Weather conditions varied over the eight nights of observations, with $\tau=0.07-0.28$.  All observations were taken at elevations between 40-70$^\circ$ where the gain curve is relatively flat. On source integration times 
ranged from $5-100$ minutes and were determined by estimating $L_{\rm CO}^\prime$ from $L_{\rm IR}$ using the average ratio from \citet{carilli2013}. 
The integration times were estimated to obtain a $>3\sigma$ detection of the expected integrated line luminosity in each source according to the LMT integration time calculator. We obtained $>3\sigma$ detections of 17 out of 24 targets. With the sole exception of 5MUSES-313, where the requested on source integration time was not completed, the desired rms was reached for all sources. The rms has been calculated using the full spectrum from 73-111\,GHz. We list the rms in Table \ref{properties} for each source. There is no correlation between rms and S/N, so low CO(1-0) emission appears to be an intrinsic property of our undetected sources.

Data were reduced and calibrated using DREAMPY (Data REduction and Analysis Methods in PYthon). DREAMPY is written by G. Narayanan and is used specifically to reduce and analyze LMT/RSR data. It is a complete data reduction package with interactive graphics. For each observation scan, four distinct spectra are produced from the RSR. After applying appropriate instrumental calibrations from the DREAMPY pipeline, certain frequencies where known instrument artifacts are sometimes present were removed. There are no known bandpass features in the spectral regions where the CO(1-0) lines are expected. For each observation, linear baselines were calculated outside the region of the CO(1-0) line, and the rms estimated from the baseline of the full spectrum is used when all data for a given source are averaged together to produce the final spectrum. The averaging is a weighted average where the weights are set to $1/{\rm rms}^2$. 

We checked whether aperture corrections were necessary using the empirical relation derived in \citet{saintonge2011a} and the optical radii from NASA/IPAC Extragalactic Database. We calculated at most a 10\% aperture correction for our lowest redshift galaxies which have the largest angular extent. Given the uncertainty inherent in applying a correction formula, we have opted not to apply any aperture corrections.

From the final spectrum for each source, in units of antenna temperature, we determined the locations of the CO(1-0) peak by fitting a simple Gaussian to the spectrum
close to the frequency $115.271/(1+z_{\rm IR})$, allowing for a generous 10\% uncertainty in $z_{\rm IR}$, where $z_{\rm IR}$ was
derived by fitting the mid-IR spectral features (described in Section \ref{sec:mir}). 
In Table \ref{properties}, we list $z_{\rm CO}$, the redshift determined from the peak of the CO(1-0) emission. 
In all cases, the redshifts derived from fitting the CO(1-0) line agree with the mid-IR redshifts within $\frac{\Delta z}{1+z_{\rm IR}}<0.1\%$. 
The RSR is designed for blind redshift searches, and we test the reliability of our CO(1-0) detections by performing blind line searches over the full 38\,GHz spectra. For 94\% (16/17) of the objects that we claim detections for, a blind line search identifies the correct redshift. For 5MUSES-225, known noise artifacts elsewhere in the spectrum confuse the redshift solution. However, since redshifts are known for our targets, we can nevertheless reliably measure the CO(1-0) emission at the correct redshift. 

Spectra (with $T_A$ in mK)
and the best fit Gaussians are
shown in Figure \ref{fig:co}. 
We integrate under the Gaussian to determine the line intensities, $I_{\rm CO}$, in K km/s. We convert the line intensity to $S_{\rm CO}\,\Delta v$ using a conversion factor of 7\,Jy/K, calibrated specifically for the LMT Early Science results, and we convert to $L^\prime_{\rm CO}$ following the equation in \citet{solomon2005} after correcting for cosmology differences. The CO(1-0) luminosities are listed in Table \ref{properties}. 

\begin{figure*}
\plotone{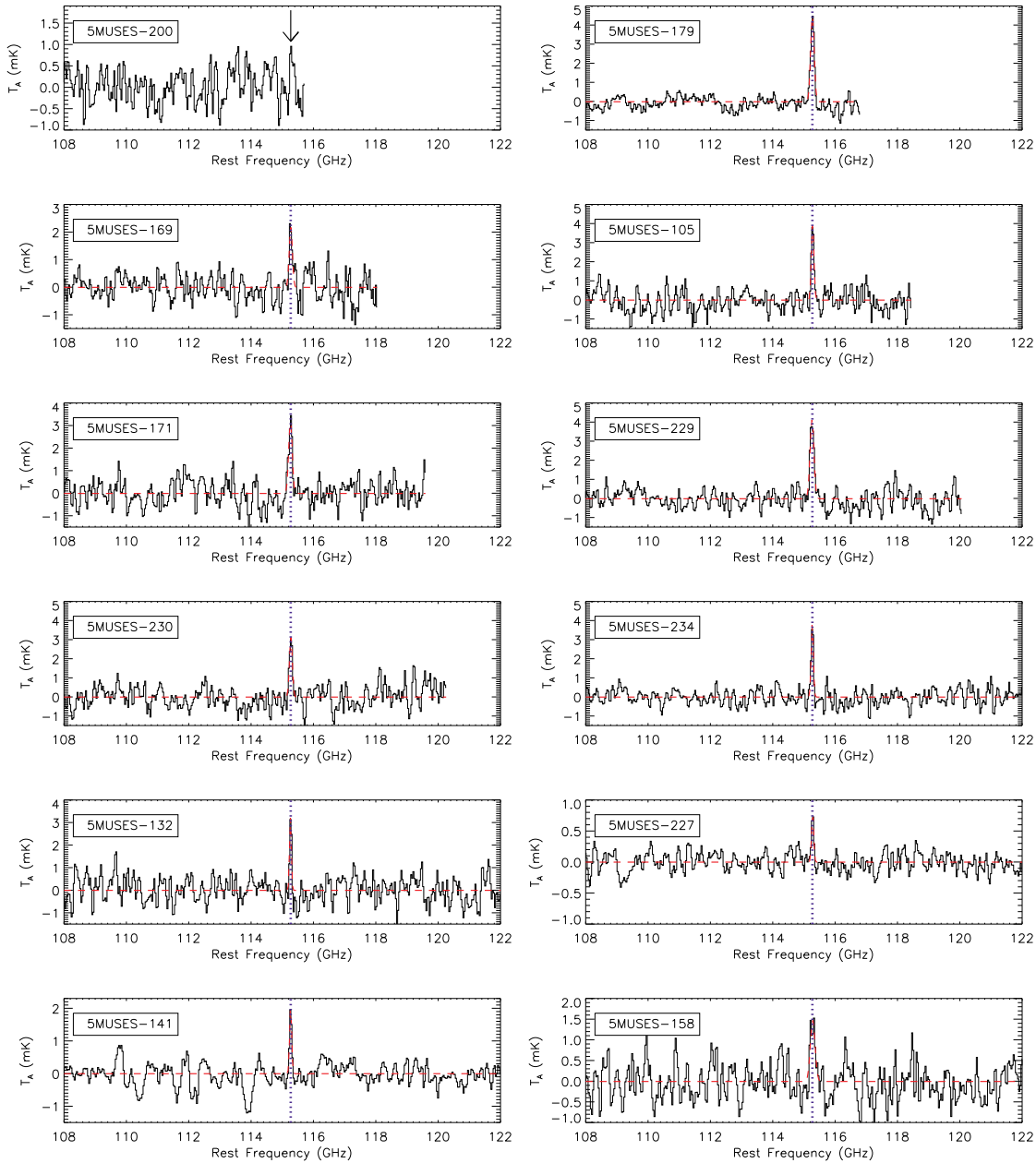}
\caption{LMT/RSR spectra for our 5MUSES subsample, with antenna temperature (mK) as a function of frequency (GHz). Here we show 14\,GHz of the rest frame spectra around the CO(1-0) line (this is only a portion of the full 73-111\,GHz observed with the RSR). We overplot the location of the peak of the CO(1-0) emission as the blue dotted line and the best fit Gaussian as the red dashed line for the 17 sources with a $3\sigma$ detection. Seven sources lack a $3\sigma$ detection, and for those, we show an arrow at the expected location of the CO(1-0) peak emission based on $z_{\rm IR}$. The spectra are presented in order of increasing redshift. \label{fig:co}}
\end{figure*}

\setcounter{figure}{0}
\begin{figure*}
\plotone{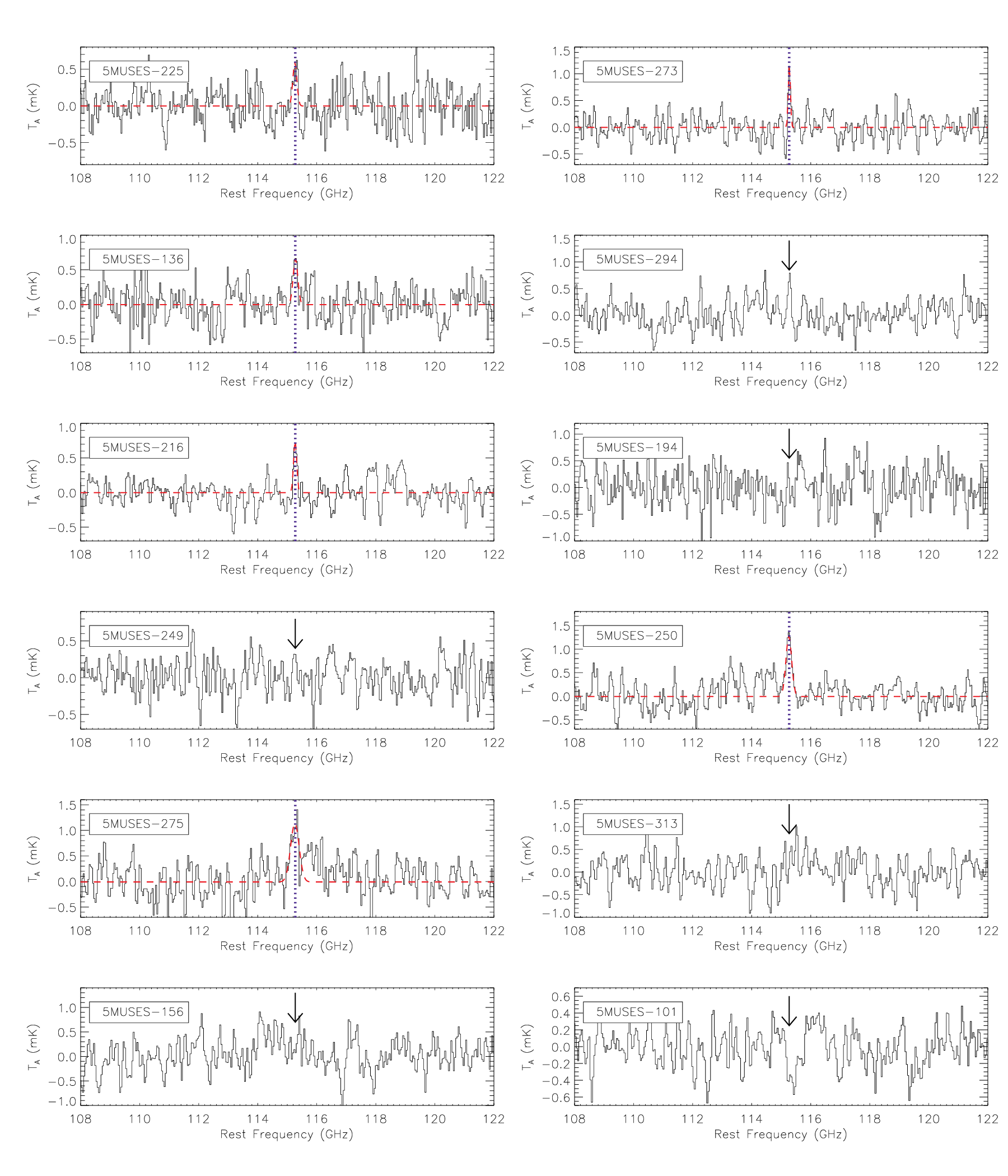}
\caption{{\it (continued)} }
\end{figure*}

\subsection{Infrared Emission}
\label{sec:mir}
Our analysis centers around comparing the molecular gas emission, traced by $L_{\rm CO}^\prime$, with the dust emission, traced primarily through $L_{\rm IR}$.
$L_{\rm IR}(5-1000\,\mu$m) values are calculated in \citet{wu2010}. The authors create synthetic IRAC photometry using the {\it Spitzer} IRS spectra and then fit a library of templates to the 
synthetic IRAC photometry and MIPS 24, 70, and 160\,$\mu
$m photometry. The final spectral energy distribution (SED) for each galaxy is created by combining the mid-IR spectroscopy with the best-fit template. Finally, the authors calculate $L_{\rm IR}$ by integrating under the SED
from $5-1000\,\mu$m. Twelve of the sources in our study also have photometry from SPIRE on the {\it Herschel Space Observatory} \citep{magdis2013}. We test whether including this longer wavelength data changes $L_{\rm IR}$ for these 12 sources, and find excellent agreement, such that there is less than a 3\% difference between $L_{\rm IR}$ calculated with and without SPIRE data. For the purposes of this study, we wish to use the standard definition: $L_{\rm IR}(8-1000\,\mu$m). We correct the $L_{\rm IR}(5-1000\,\mu$m) values from \citet{wu2010} to 
$L_{\rm IR}(8-1000\,\mu$m) by scaling by 0.948, a conversion factor which was determined using composite LIRG and ULIRG SEDs from \citet{kirkpatrick2012}. 
In addition, \citet{wu2010} uses a slightly different cosmology with 
$H_0=70\,$km\,s$^{-1}$\,Mpc$^{-1}$. Over our range of redshifts, this requires an average conversion factor of 0.970 to account for the difference in cosmologies. To summarize, we calculate:
\begin{equation}
L_{\rm IR}^{\rm tot}(8-1000\,\mu{\rm m})=L_{\rm IR}^{\rm Wu} \times 0.948 \times 0.970
\end{equation} 
These values are listed in Table \ref{tbl:dust}.

\begin{figure}
\plotone{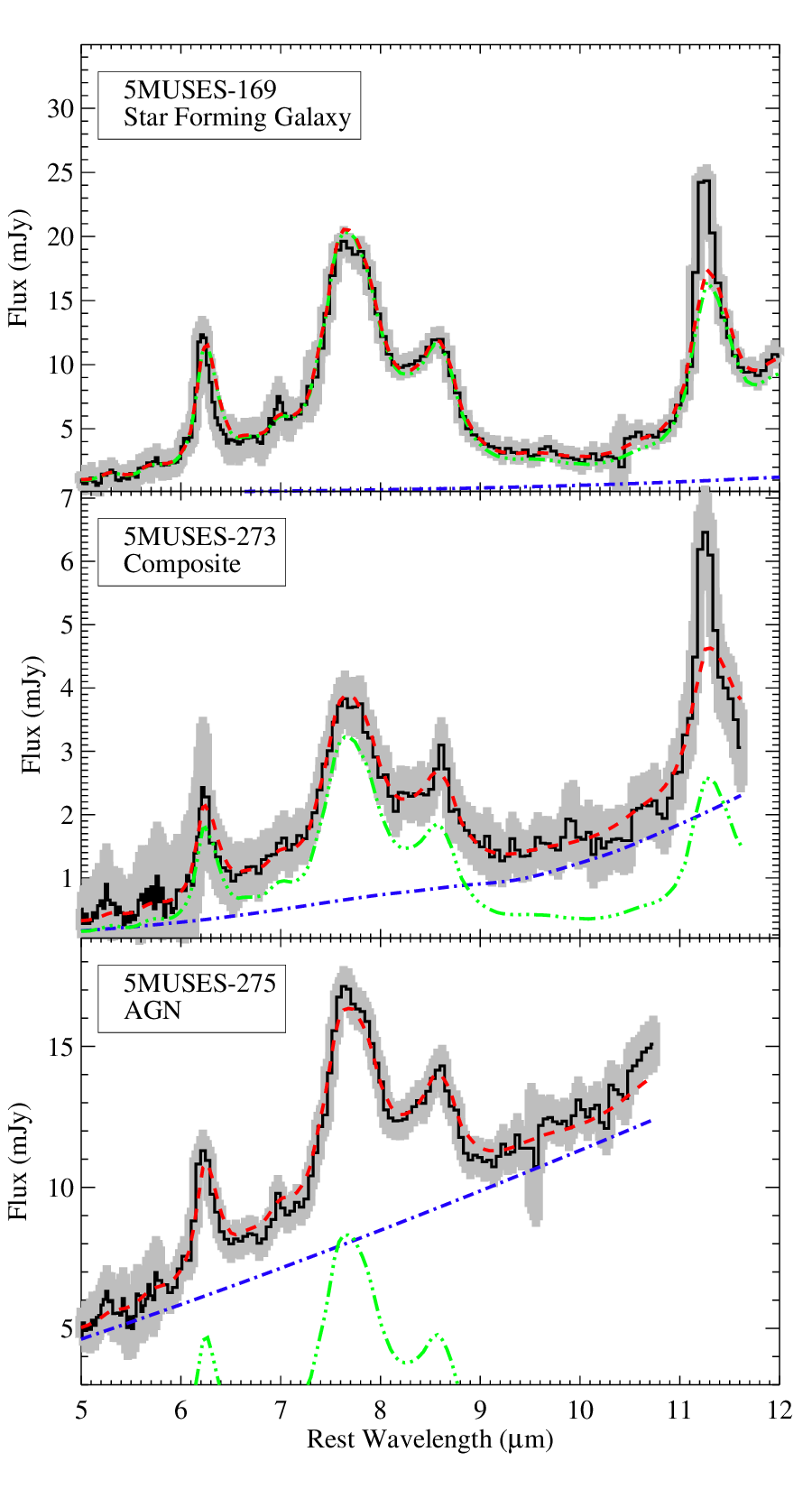}
\caption{We determine the presence and strength of a mid-IR AGN by decomposing the mid-IR spectrum into a star formation component (green triple-dot-dashed line) and a power-law continuum with extinction (blue dot-dashed line). We calculate $f_{\rm AGN,midIR}$ as the fraction of the luminosity of the best fit model (red dashed line) due to the power-law component. We illustrate the spectral decomposition of a star forming galaxy (top panel), composite (middle panel), and AGN (bottom panel).\label{fig:decomp}}
\end{figure}

When an AGN is present, $L_{\rm IR}^{\rm tot}$ can have a non-neglible
contribution from dust heated by an AGN, and hence $L_{\rm IR}$ does not directly translate to an SFR. 
We diagnose the presence and strength of an AGN through mid-IR ($5-15\,\mu$m rest frame) spectral decomposition.
We follow the technique outlined in detail in \citet{pope2008} which we summarize here. We fit the individual spectra with a
model comprised of three components: (1) the star formation component is represented by the mid-IR spectrum of the prototypical starburst M~82 \citep{schreiber2003}. We have verified our choice of starburst template by comparing with the starburst composite SED from \citet{brandl2006} and find no difference in the decomposition results;
(2) the AGN component is determined by fitting a pure power-law with the slope and normalization
as free parameters; (3) an extinction curve from the \citet{draine2003} dust models is applied to the AGN component. The extinction curve is not monotonic in wavelength
and contains silicate absorption features, the most notable for our wavelength range being at 9.7$\,\mu$m. We tested applying additional extinction to the star formation component 
beyond that inherent in the M~82 template and found this to be negligible for all sources. We fit all three components simultaneously.
For each source, we quantify the strength of the mid-IR AGN, $f_{\rm AGN,midIR}$, as the fraction of the total mid-IR luminosity coming from the power-law 
continuum component. The PAH EW from \citet{wu2010}, initially used to select our sample, directly relates to the mid-IR AGN fraction, supporting the reliability of our diagnostic. Our decomposition technique is illustrated in Figure \ref{fig:decomp}.

We use the mid-IR AGN strength, $f_{\rm AGN,midIR}$, to determine the total contribution of the AGN to $L_{\rm IR}^{\rm tot}$. To estimate the conversion between
$f_{\rm AGN,midIR}$ and $f_{\rm AGN,IR}$, we use 22 composite SEDs created from $>300$ LIRGs and ULIRGs spanning a redshift range of $z=0.5-3$ (Kirkpatrick et al.\,2015, in preparation). We decompose the mid-IR portion of the composite SEDs following the technique outlined above. We then decompose the full SED, from 0.5-1000\,$\mu$m, by fitting a pure AGN template and a star formation template.
We find that the total IR AGN contribution is $53\pm12\%$ of the mid-IR AGN strength ($f_{\rm AGN,IR}=0.53\times f_{\rm AGN,midIR}$).
For the 5MUSES sample, where we lack enough data points to accurately decompose the full-IR SED, 
we scale the mid-IR AGN strength by 0.53 to estimate the total AGN contribution, $f_{\rm AGN,IR}$, to $L_{\rm IR}^{\rm tot}$, and we use this value to calculate $L_{\rm IR}^{\rm SF}$, the IR luminosity attributed to star formation.
In the analysis that follows, we separate our sources according to $f_{\rm AGN,midIR}$, resulting in three
categories: (1) purely star forming galaxies have $f_{\rm AGN,midIR}<0.2$ ($f_{\rm AGN,IR}<0.1$); (2) AGN have $f_{\rm AGN,midIR}>0.7$ ($f_{\rm AGN,IR}>0.4$); and (3) composites have $f_{\rm AGN,midIR}=0.2-0.7$ ($f_{\rm AGN,IR}=0.1-0.4$). 

We are also interested in the strength of the PAH emission, since this is expected to trace star formation and photodissociation regions within molecular clouds. By comparing the PAH emission with $L_{\rm IR}$ and CO(1-0) emission, we have three separate tracers of the dust and gas in the ISM which are expected to correlate with star formation. We quantify the PAH emission as $L_{6.2}$, the luminosity of the isolated 6.2\,$\mu$m line.
We fit a continuum on either side of this feature at 5.9\,$\mu$m and 6.5\,$\mu$m and remove the continuum component. We then integrate under the continuum subtracted emission feature to obtain $L_{6.2}$. None of our sources are extended at 8.0\,$\mu$m using the {\it Spitzer} IRAC images, so aperture corrections are not necessary when comparing the {\it Spitzer} and LMT luminosities. We list $L_{6.2}$, $L_{\rm IR}^{\rm SF}$, $f_{\rm AGN,midIR}$, and $f_{\rm AGN,IR}$ in Table \ref{tbl:dust}.

\begin{deluxetable*}{lc cccc cccc}
\tablecolumns{10}
\tablecaption{Infrared and Star Formation Properties \label{tbl:dust}}
\tablehead{\colhead{ID} & \colhead{$\log L_{\rm IR}^{\rm tot}$\tablenotemark{a}} & \colhead{$f_{\rm AGN,midIR}$} &
\colhead{$f_{\rm AGN,IR}$\tablenotemark{b}} & \colhead{Type\tablenotemark{c}} & \colhead{$\log L_{\rm IR}^{\rm SF}$} & 
\colhead{$\log L_{6.2}$} & \colhead{$\log M_\ast$} & \colhead{sSFR\tablenotemark{d}} & \colhead{Desig.\tablenotemark{e}}\\
\colhead{} & \colhead{($L_\odot$)} & \colhead{} & \colhead{} & \colhead{} & \colhead{($L_\odot$)} & \colhead{($L_\odot$)} & \colhead{($M_\odot$)} & \colhead{(Gyr$^{-1}$)} & \colhead{}}
\startdata
5MUSES-200	& 10.36	& 0.23	& 0.12	& Comp & 10.31	& 8.04 	& 10.48	& 0.09	& MS \\
5MUSES-179	& 10.22	& 0.24	& 0.13	& Comp & 10.17	& 7.95 	& 11.08	& 0.01	& MS \\
5MUSES-169	& 10.79	& 0.06	& 0.03	& SFG 	& 10.79	& 8.45 	& 10.82	& 0.11	& MS \\
5MUSES-105	& 10.88	& 0.16	& 0.08 	& SFG 	& 10.85	& 8.38 	& 10.79	& 0.13	& MS \\
5MUSES-171	& 11.06	& 0.08	& 0.04	& SFG	& 11.05	& 8.54 	& 10.93	& 0.16	& MS \\
5MUSES-229	& 11.10	& 0.12	& 0.06	& SFG	& 11.08	& 8.62 	& 11.36	& 0.06	& MS \\
5MUSES-230	& 11.09	& 0.00	& 0.00	& SFG	& 11.09	& 8.98 	& 10.46	& 0.50	& SB \\
5MUSES-234	& 11.03	& 0.15	& 0.08	& SFG	& 11.00	& 8.52 	& 10.47	& 0.40	& SB \\
5MUSES-132	& 11.30	& 0.00	& 0.00 	& SFG	& 11.30	& 9.16 	& 10.67	& 0.51	& SB \\
5MUSES-227	& 11.08	& 0.57	& 0.30 	& Comp	& 10.91	& 8.63 	& 10.73	& 0.18	& MS \\
5MUSES-141	& 11.14	& 0.67	& 0.36	& Comp	& 10.93	& 8.71 	& 11.15	& 0.07	& MS \\
5MUSES-158	& 11.41	& 0.23	& 0.12	& Comp	& 11.36	& 8.87 	& 10.99	& 0.28	& SB \\
5MUSES-225	& 11.09	& 0.29	& 0.15	& Comp	& 11.02	& 8.61 	& 10.99	& 0.13	& MS \\
5MUSES-273	& 11.40	& 0.45	& 0.24	& Comp	& 11.28	& 8.78 	& 10.97	& 0.23	& MS \\
5MUSES-136	& 11.39	& 0.77	& 0.41	& AGN	& 11.13	& 9.03 	& 11.41	& 0.06	& MS \\
5MUSES-294	& 11.55	& 0.16	& 0.08	& SFG	& 11.52	& 8.97 	& 11.49	& 0.12	& MS \\
5MUSES-216	& 11.39	& 0.14	& 0.07	& SFG	& 11.37	& 9.06 	& 10.71	& 0.53	& SB \\
5MUSES-194	& 11.72	& 0.81	& 0.43	& AGN	& 11.44	& 8.34 	& 11.19	& 0.21	& MS \\
5MUSES-249	& 11.63	& 0.12	& 0.06	& SFG	& 11.61	& 9.26 	& 10.96	& 0.52	& SB \\
5MUSES-250 	& 11.63	& 0.05	& 0.03	& SFG	& 11.63	& 9.44 	& 11.29	& 0.26	& MS \\
5MUSES-275	& 12.00	& 0.79	& 0.42	& AGN	& 11.73	& 9.46 	& 11.13	& 0.47	& SB \\
5MUSES-313	& 11.81	& 0.41	& 0.22	& Comp	& 11.70	& 9.12	& 10.64	& 1.36	& SB \\
5MUSES-156	& 12.06	& 0.31	& 0.16	& Comp	& 11.99	& 9.24 	& 10.17	& 7.70	& SB \\
5MUSES-101	& 11.91 	& 0.82	& 0.43 	& AGN	& 11.63	& 9.03	& 10.86	& 0.68	& SB
\enddata
\tablenotetext{a}{The $L_{\rm IR}^{\rm tot}$ values are scaled from \citet{wu2010}; see Section \ref{sec:mir} for details.}
\tablenotetext{b}{The fraction of $L_{\rm IR}(8-1000\,\mu$m) due to heating by an AGN.}
\tablenotetext{c}{Star forming galaxies (SFG) have $f_{\rm AGN,midIR}<0.2$ ($f_{\rm AGN,IR}<0.1$); AGN have $f_{\rm AGN,midIR>0.7}$ ($f_{\rm AGN,IR}>0.4$); composites (Comp) have $f_{\rm AGN,midIR}=0.2-0.7$ ($f_{\rm AGN,IR}=0.1-0.4$)}. 
\tablenotetext{d}{Calculated using $L_{\rm IR}^{\rm SF}$.}
\tablenotetext{e}{Main Sequence (MS) or Starburst (SB) according to Equation \ref{eq:MS}, where sSFR is calculated using $L_{\rm IR}^{\rm SF}$.}
\end{deluxetable*}

\subsection{Specific Star Formation Rates}
The specific star formation rate (sSFR) is the ratio of SFR to stellar mass.
\citet{shi2011} determined stellar masses for the 5MUSES sample by fitting \citet{bruzual2003} population synthesis models to optical and near-IR broadband photometry assuming a Chabrier IMF. 
We adopt these stellar masses, and we calculate the SFR from $L_{\rm IR}^{\rm SF}$ according to
\begin{equation}
\left(\frac{\rm SFR}{M_\odot\, {\rm yr}^{-1}}\right) = 1.509\times 10^{-10} \left(\frac{L_{\rm IR}^{\rm SF}}{L_\odot}\right) 
\end{equation}
assuming a Kroupa IMF and a constant star formation rate over the past 100\,Myr \citep{murphy2011}. We do not convert between a Chabrier IMF and a Kroupa IMF for this formula, since the conversion is very small \citep{zahid2012}.

\begin{figure}
\plotone{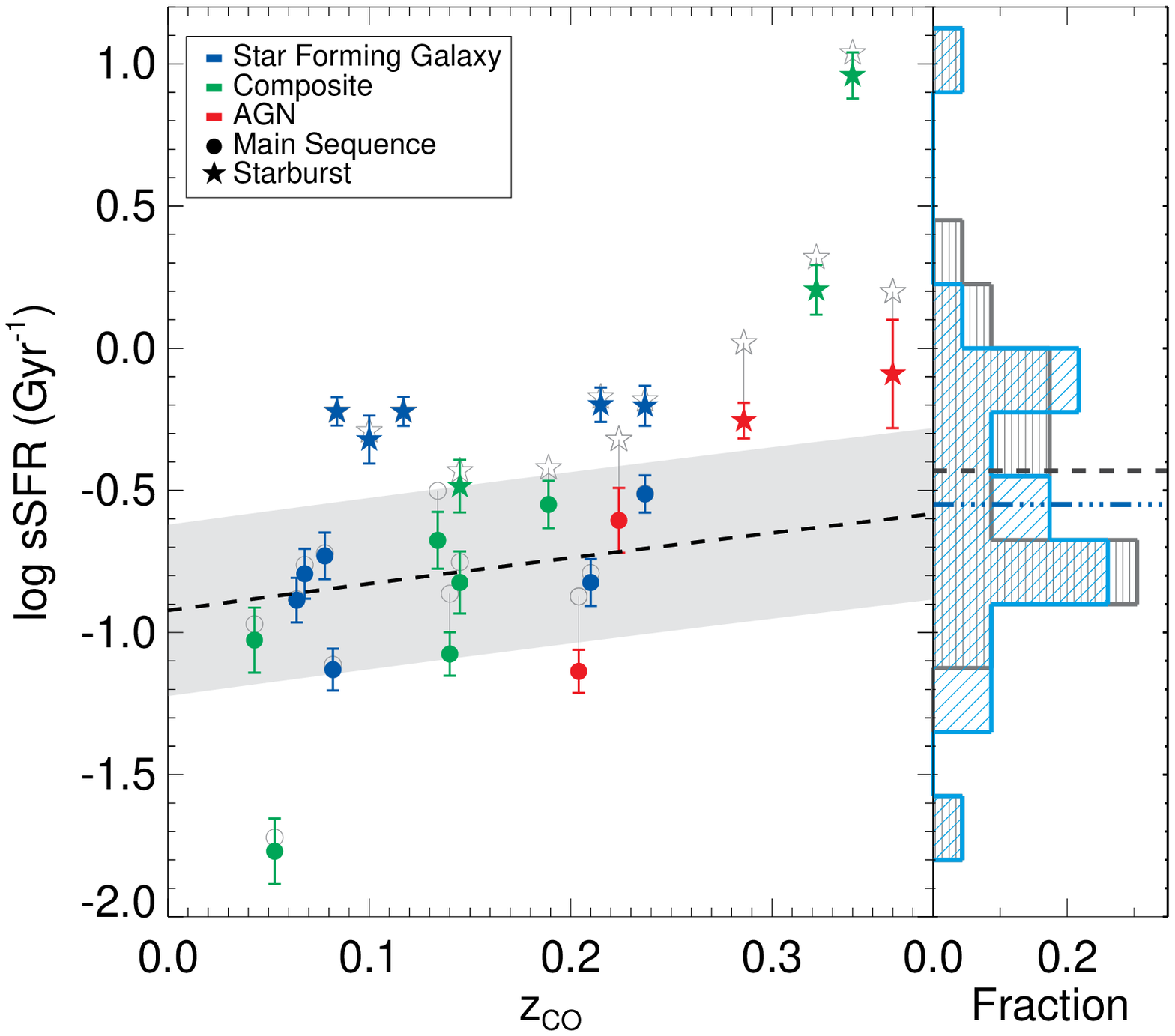}
\caption{sSFR vs. $z$ for our sample. The open symbols show sSFR calculated with $L_{\rm IR}^{\rm tot}$ while the filled symbols show sSFR calculated with $L_{\rm IR}^{\rm SF}$. We color the sources according to whether they are star forming galaxies (blue), composites (green), or AGN (red), according to $f_{\rm AGN,midIR}$ (star forming galaxies have $f_{\rm AGN,midIR}<0.2$; AGN have $f_{\rm AGN,midIR}>0.7$; and composites have $f_{\rm AGN,midIR}=0.2-0.7$). We overplot the main sequence relation from Equation \ref{eq:MS} (dashed line), and the grey shaded region extends a factor of two above and below this line, consistent with the scatter measured in \citet{elbaz2011}. Removing the AGN contribution has the effect of generally lowering the sSFR and  moving two sources onto the main sequence. We show the distributions of sSFRs in the histogram on the right. The grey histogram is the distribution when sSFR is calculated with $L_{\rm IR}^{\rm tot}$, and the cyan histogram is the distribution of sSFR calculated with $L_{\rm IR}^{\rm SF}$. The dashed and dot-dashed lines show the medians.\label{fig:MS}}
\end{figure}

The galaxy main sequence can be used to classify galaxies as either normal star forming galaxies or starbursts based on whether they have an enhanced SFR for a given $M_\ast$.
This relationship is a slowly varying function of redshift.
\citet{elbaz2011} present a relationship between sSFR and the time since the Big Bang in Gyr, $t_{\rm cosmic}$, for the galaxy main sequence \citep[see Equation 13 in][]{elbaz2011}.
The relation in \citet{elbaz2011} is derived assuming a Salpeter IMF, an $L_{\rm IR}$-SFR conversion from \citet{kennicutt1998}, and a cosmology with $H_0=70\,$km\,s$^{-1}$\,Mpc$^{-1}$, $\Omega_m=0.3$, and $\Omega_\Lambda=0.7$. We convert the relation from a Salpeter IMF to a Kroupa IMF using $M_\ast^{\rm Kroupa}=0.62\,M_\ast^{\rm Salpeter}$ \citep{zahid2012}. We convert $t_{\rm cosmic}$ to the cosmology used in this paper by multiplying by 1.02, appropriate for our redshift range. Finally, the SFRs are related by 
SFR$_{\rm Murphy11} = 0.86\,{\rm SFR}_{\rm Kennicutt98}$ \citep{kennicutt2012}. Applying all of these conversions gives the MS relation appropriate for the present work:
\begin{equation}
\label{eq:MS}
{\rm sSFR}_{\rm MS}\, ({\rm Gyr^{-1}}) =  38\times t_{\rm cosmic}^{-2.2}
\end{equation}
If a galaxy has a sSFR a factor of two greater than sSFR$_{\rm MS}$, it is classified as a starburst. We use $z_{\rm CO}$ and $L_{\rm IR}^{\rm SF}$ to calculate sSFR$_{\rm MS}$ and list 
the stellar masses along with the main sequence/starburst designations in 
Table \ref{tbl:dust}. We plot sSFR vs. $z$ for our sample in Figure \ref{fig:MS}. We show the sSFR calculated using $L_{\rm IR}^{\rm tot}$ (grey open symbols) and the sSFR calculated using $L_{\rm IR}^{\rm SF}$ (filled symbols). We color the sources according to whether they are star forming galaxies (blue), composites (green), or AGN (red), according to $f_{\rm AGN,IR}$ (see Section \ref{sec:mir}). Removing the AGN contribution to $L_{\rm IR}^{\rm tot}$, which we have done for all sources, has the effect of moving two of our sources from the starburst region onto the main sequence. Two sources, 5MUSES-136 and 5MUSES-179, lie below the main sequence, possibly indicating that they are transitioning to a more quiescent evolutionary stage. In this small sample, we observe no obvious association of AGN strength with distance from the main sequence, similar to what is seen in \citet{elbaz2011}.

\section{Results \& Discussion}
\subsection{Relationship Between Molecular Gas and Dust Emission}
In the local Universe, the SK law is traditionally expressed in terms of surface densities, which necessarily require resolved measurements of star formation and molecular gas \citep{schmidt1959,kennicutt1998,shi2011,kennicutt2012}. At higher redshifts, resolved measurements
are often not feasible, and global measurements of the SFR and molecular gas are used instead.
The ratio of $L_{\rm IR}$, directly related to a global SFR, and $L^\prime_{\rm CO}$ is commonly referred to as a star formation efficiency \citep{young1991,solomon2005}.
It is essentially the integrated version of the SK law without the uncertain conversion from CO to H$_2$ mass. This quantity is also related to the inverse of the gas depletion timescale which describes how long a galaxy could continue to form stars at the current rate if the gas reservoir is not replenished \citep[e.g.,][]{saintonge2011}.

In general, $L_{\rm IR}$ correlates with $L^\prime_{\rm CO}$, although there is significant scatter. $L_{\rm IR}$ is primarily measuring the reradiated light from newly formed stars (with some contribution from an older stellar population), while $L^\prime_{\rm CO}$ is measuring the reservoir of molecular gas available to form stars in the future; hence 
the dust and gas do not necessarily trace star formation on the same timescales.
CO emission can also be present between molecular clouds, introducing more scatter in the relationship between $L_{\rm IR}$ and $L_{\rm CO}^\prime$. To account for this scatter, many authors propose two relationships, one for starbursts, undergoing an enhanced $L_{\rm IR}/L_{\rm CO}^\prime$, and one for normal star forming galaxies \citep[e.g.,][]{daddi2010,genzel2010,carilli2013,tacconi2013}.
 If a galaxy undergoes a burst of star formation that also triggers AGN growth, this could account for some of the scatter in $L_{\rm IR}/L_{\rm CO}^\prime$.
Initially, as an embedded AGN grows more luminous, it heats some of the surrounding dust, but is enshrouded enough that the host
galaxy is still visible \citep[e.g.,][]{sanders1988,hopkins2006,kirkpatrick2012}. 
The increase in the amount of warm dust heated by the AGN will enhance $L_{\rm IR}$ but may not yet affect $L^\prime_{\rm CO}$, 
implying an artificially high $L_{\rm IR}/L_{\rm CO}^\prime$ unless the AGN contribution is accounted for (e.g., by considering $L_{\rm IR}^{\rm SF}$).
As the AGN becomes less enshrouded, the PAH and cold dust emission from the host become less prominent because the dust heated by the AGN is outshining the dust in star forming 
regions, and/or because feedback 
from the AGN is quenching the star formation. If the AGN is quenching the star formation, this will produce lower $L_{\rm IR}^{\rm SF}$ values and hence lower $L_{\rm IR}^{\rm SF}/L_{\rm CO}^\prime$, unless feedback from the 
AGN expels CO on the same timescales.

\begin{figure}
\plotone{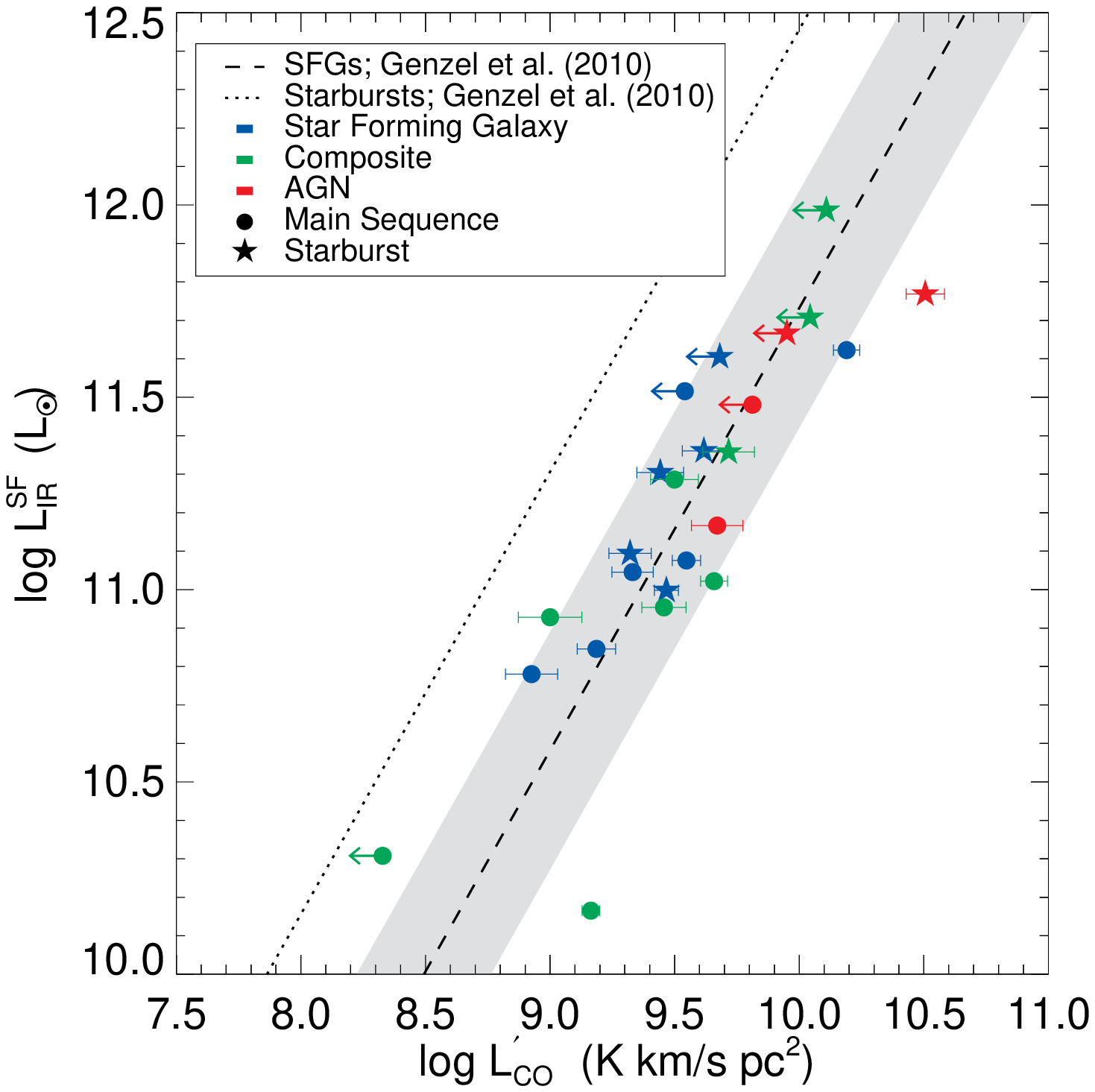}
\caption{The relationship between $L_{\rm IR}^{\rm SF}$ and $L^\prime_{\rm CO}$ for the 5MUSES sample. We also include the relationships for $L_{\rm IR}$ and $L^\prime_{\rm CO}$ derived for starbursts and star forming galaxies (SFGs) from \citet{genzel2010}, where the grey shaded region indicates the standard deviation. We color the points according to mid-IR power source, and we use different symbols to indicate the galaxies that are starbursting according to the sSFR criterion. There is no strong separation according to either mid-IR power source or starburstiness, and our galaxies all lie close to the SFG relation from \citet{genzel2010}. \label{fig:lir_lco}}
\end{figure}

As yet, no study has attempted to quantify the effect of the AGN on $L_{\rm IR}/L_{\rm CO}^\prime$ individually in galaxies due to the difficulty in measuring the amount of $L_{\rm IR}$ attributable to an AGN. 
The 5MUSES sample has mid-IR spectra that exhibit both PAH features and a strong underlying continuum, allowing us to cleanly separate out the AGN luminosity 
and calculate $L_{\rm IR}^{\rm SF}$. Figure \ref{fig:lir_lco} shows the relationship between $L_{\rm IR}^{\rm SF}$ and $L^\prime_{\rm CO}$, and the points are colored according to $f_{\rm AGN,IR}$. We overplot the relations between $L_{\rm IR}^{\rm SF}$ and $L^\prime_{\rm CO}$ determined for star forming galaxies (SFGs, dashed line; grey region indicates one standard deviation) and starbursts (dotted line) in
\citet{genzel2010}. These relations were robustly determined using large samples of normal star forming galaxies and starbursts from $z\sim0-3$, and we are interested in testing how removing the AGN component in our galaxies changes their position relative to these relations. The presence of two separate relations is discussed in depth in \citet{carilli2013}, where the authors also present a single relation fitting all available $L_{\rm IR}$ and $L_{\rm CO}$ measurements in the literature. We compare to the separate starburst and SFG relations to investigate any difference in $L_{\rm IR}^{\rm SF}/L_{\rm CO}^\prime$ according to IR power source or starburstiness.

\citet{genzel2010} determined the relations for SFGs and starbursts using the far-IR luminosity (FIR: $50-300\,\mu$m) as opposed to $L_{\rm IR}$, so we scale the relationships according to $L_{\rm IR}/{\rm FIR}=1.63$, a ratio determined using the LIRG and ULIRG templates from \citet{kirkpatrick2012}. We overplot the $1\sigma$ scatter from \citet{genzel2010} as the grey shaded region.
We plot the main sequence galaxies (determined by Equation \ref{eq:MS}) as filled circles and starbursts as filled stars. The starbursts on average have a factor of two 
higher $L_{\rm IR}^{\rm SF}$ \underline{and} $L^\prime_{\rm CO}$ than the main sequence galaxies. 
We find that both the main sequence and starbursts are consistent with the SFG relationship from \citet{genzel2010}, and there is no strong separation according to IR power source.

\begin{figure}
\plotone{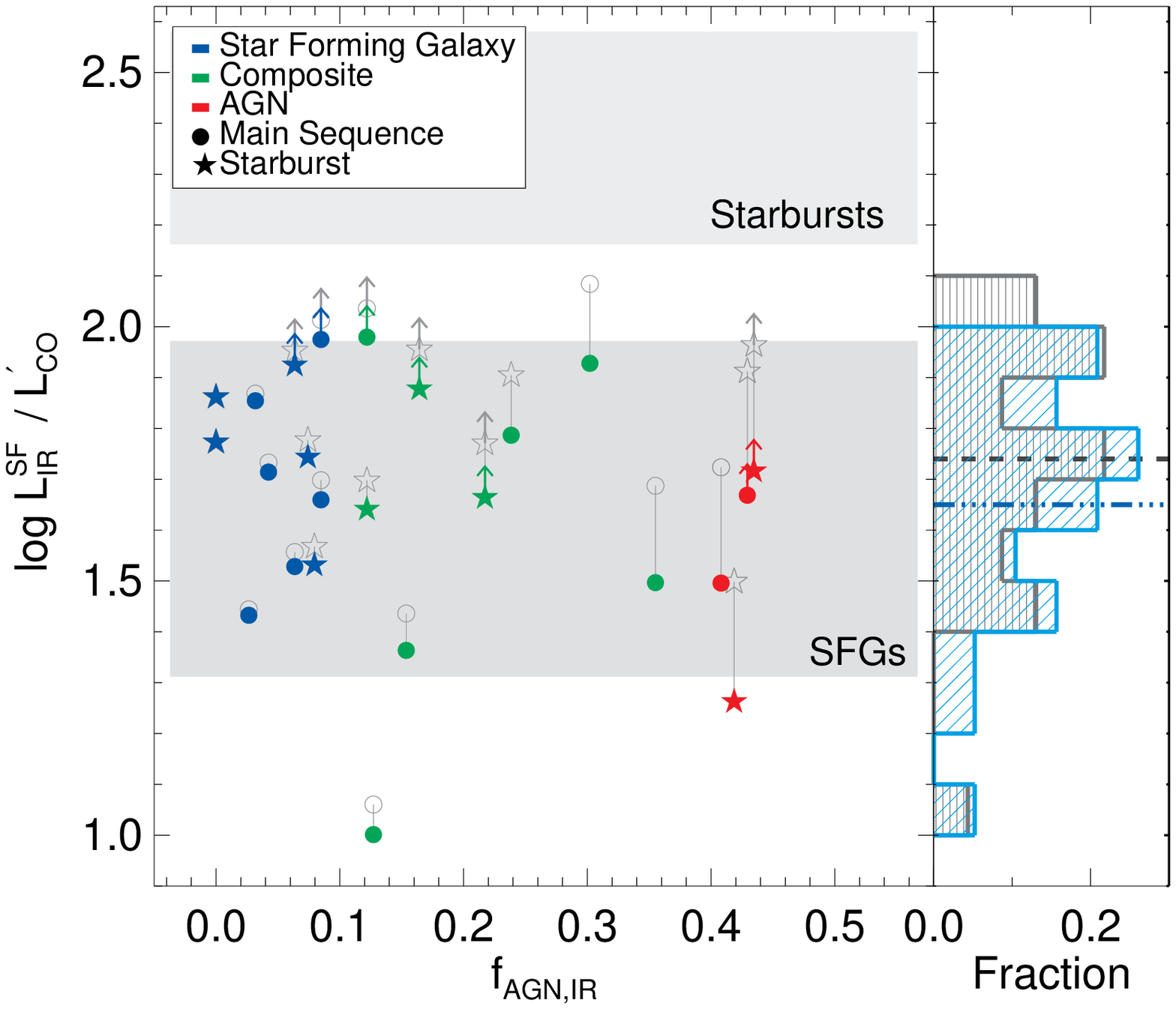}
\caption{$L_{\rm IR}^{\rm SF}/L_{\rm CO}^\prime$ vs. $f_{\rm AGN,IR}$. Unfilled symbols are calculated using $L_{\rm IR}^{\rm tot}$, uncorrected for AGN contamination. Symbols correspond to main sequence/starburst designation according to sSFR. The shaded regions show one standard deviation around the mean $L_{\rm IR}/L_{\rm CO}^\prime$ for SFGs and starbursts from \citet{genzel2010}. Nearly all galaxies lie within the SFG region, and after removing the AGN contribution to $L_{\rm IR}^{\rm tot}$, we calculate a mean $L_{\rm IR}^{\rm SF}/L_{\rm CO}^\prime$ consistent with the mean calculated for the larger sample in \citep{genzel2010}. There is no significant trend with $f_{\rm AGN,IR}$, although the AGN do have a lower average $L_{\rm IR}^{\rm SF}/L_{\rm CO}^\prime$ than the star forming galaxies or the composites. We show the distributions in the histogram on the right. The grey histogram is the distribution of $L_{\rm IR}^{\rm tot}/L_{\rm CO}^\prime$, with the mean overplotted as the dashed line, and the cyan histogram is the distribution of $L_{\rm IR}^{\rm SF}/L_{\rm CO}^\prime$, with the dot-dashed line illustrating the mean. \label{fig:SFE}}
\end{figure}

We can look at this more simply by considering the ratio $L_{\rm IR}^{\rm SF}/L^\prime_{\rm CO}$. In Figure \ref{fig:SFE}, we plot $L_{\rm IR}^{\rm SF}/L^\prime_{\rm CO}$ as a function of $f_{\rm AGN,IR}$. We also include $L_{\rm IR}^{\rm tot}/L_{\rm CO}^\prime$ as the unfilled
circles and stars. The grey shaded regions illustrate the standard deviation around the average $L_{\rm IR}/L_{\rm CO}^\prime$ calculated for starbursts and SFGs in \citet{genzel2010}. The standard deviation for the SFG region is 0.33 dex, slightly larger than the standard deviation illustrated in Figure \ref{fig:lir_lco}. This is due to the fact that the relation between $L_{\rm IR}$ and $L_{\rm CO}^\prime$ is non-linear, so the standard deviation relative to the mean $L_{\rm IR}/L_{\rm CO}^\prime$ is larger.


None of our galaxies lie in the starburst region.
\citet{genzel2010} find a mean $L_{\rm IR}/L^\prime_{\rm CO}$ of 44. We calculate that the mean $L_{\rm IR}^{\rm tot}/L_{\rm CO}^\prime$ is 59, but when we remove the AGN contribution to $L_{\rm IR}^{\rm tot}$, we calculate a mean $L_{\rm IR}^{\rm SF}/L^\prime_{\rm CO}$ of 52, closer to the mean measured by \citet{genzel2010}. Removing the AGN component only mildly reduces the scatter about the mean $L_{\rm IR}^{\rm SF}/L^\prime_{\rm CO}$.
Two galaxies, 5MUSES-179 and 5MUSES-275, lie below the SFG region. The galaxy with the lowest $L_{\rm IR}^{\rm SF}/L_{\rm CO}^\prime$, 5MUSES-179, also lies decidedly below the main sequence region in Figure \ref{fig:MS}, further indicating that its star formation is highly inefficient, and this galaxy could be transitioning to a quiescent phase.
We find no relationship between $L_{\rm IR}^{\rm SF}/L_{\rm CO}^\prime$ and $f_{\rm AGN,IR}$. For this small sample, the star forming galaxies exhibit less scatter in $L_{\rm IR}^{\rm SF}/L_{\rm CO}^\prime$ than the composite galaxies, although both groups have the same average $L_{\rm IR}^{\rm SF}/L_{\rm CO}^\prime$. The AGN have a lower average $L_{\rm IR}^{\rm SF}/L_{\rm CO}^\prime$ indicating that these galaxies are not converting gas to stars at the same rate as the composite or SF galaxies; this hints that the
star formation might be beginning to quench in these AGN sources.

It is interesting to note that two galaxies with very high $L_{\rm IR}^{\rm SF}/L_{\rm CO}^\prime$ lower limits (5MUSES-249 and 5MUSES-294) show no trace of an AGN according to their mid-IR spectra. This could be illustrative of the different timescales that AGN signatures, starburst signatures, and enhanced $L_{\rm IR}^{\rm SF}/L_{\rm CO}^\prime$ ratios are visible. Mid-IR spectroscopy and Chandra X-ray observations provide evidence that the majority of local ULIRGs and high redshift submillimeter galaxies (SMGs), which have a high merger (and hence, starburst) fraction and enhanced $L_{\rm IR}/L_{\rm CO}^\prime$, are predominately powered by star formation in the mid-IR \citep[e.g.,][]{alexander2005,pope2008,veilleux2009,elbaz2011}. A non-negligible fraction of LIRGs and ULIRGs host a mid-IR luminous AGN, although optical morphologies of local LIRGs reveal that mid-IR AGN signatures are predominately found either in isolated disk galaxies or coalesced nuclei at the end of a merger \citep{petric2011}. We expect, then, that if the AGN and starburst phase do not overlap completely, AGN should have lower $L_{\rm IR}^{\rm SF}/L_{\rm CO}^\prime$ ratios, and Figure \ref{fig:SFE} shows that they do, for this small sample, once the AGN contribution to $L_{\rm IR}^{\rm tot}$ has been accounted for.

%

\begin{figure}
\plotone{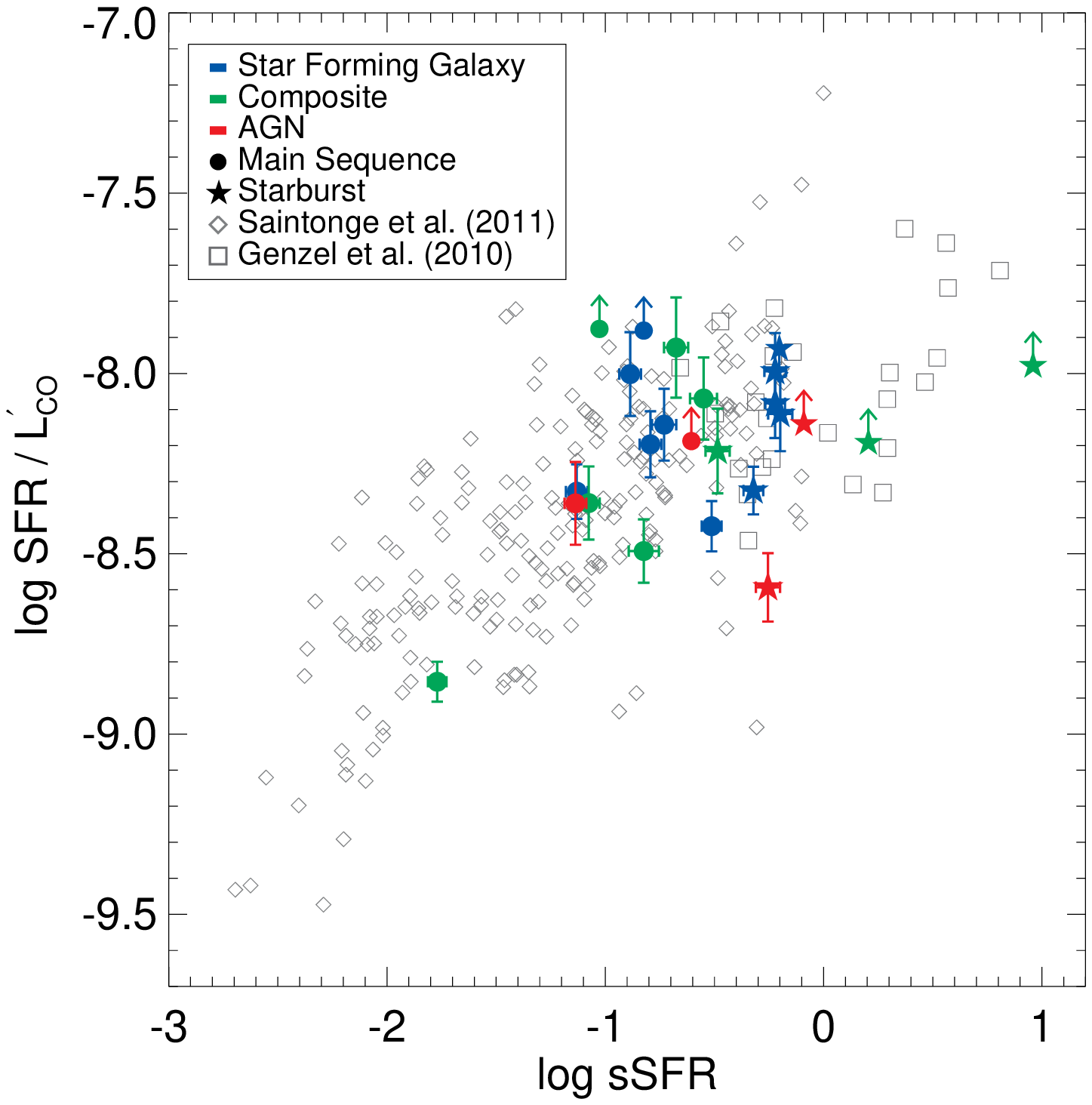}
\caption{SFR/$L_{\rm CO}^\prime$/SFR, which is related to the inverse of the gas depletion timescale, as a function of sSFR. We include the local COLD GASS sources from \citet{saintonge2011} and the $z=1-3$ sample from \citet{genzel2010}. There is a strong correlation between the two parameters for all galaxies, although this correlation would be missed if only considering the 5MUSES sample.\label{fig:tdep}}
\end{figure}

Also of interest is that we do not observe any dichotomy in $L_{\rm IR}^{\rm SF}/L_{\rm CO}^\prime$ either as a function of mid-IR power source or along the main sequence/starburst classification.
\citet{saintonge2011} observe a clear relationship between the gas depletion timescale and the sSFR in a large, complete, sample of local galaxies as part of the COLD GASS survey. We compare our sources with the COLD GASS galaxies in Figure \ref{fig:tdep}. \citet{saintonge2011} calculate the depletion timescale as $t_{\rm dep} =M_{{\rm H}_2}/{\rm SFR}$,
but to avoid any uncertainties due to converting $L_{\rm CO}^\prime$ to $M_{\rm H_2}$ (see Section \ref{sec:gas}), we simply use the ratio SFR/$L_{\rm CO}^\prime \propto t_{\rm dep}^{-1}$. We have calculated the SFRs for our galaxies using $L_{\rm IR}^{\rm SF}$. Figure \ref{fig:tdep} also includes the $z=1-3$ galaxies from \citet{genzel2010}, where we have corrected the SFRs, calculated using the conversion in \citet{kennicutt1998}, to be on the same scale as ours. The SFRs from \citet{saintonge2011} are calculated by fitting the SED from the UV out to 70\,$\mu$m. We do not further correct for differing cosmologies as the intrinsic scatter in Figure \ref{fig:tdep} is larger than any shift introduced in this manner.

When looking at just our 5MUSES sample in Figure \ref{fig:tdep}, there is no strong correlation between SFR/$L_{\rm CO}^\prime$ and sSFR, just as we observed no separation according to sSFR in Figure \ref{fig:SFE}. However, when we extend the dynamical range of the plot by considering the local galaxies from \citet{saintonge2011} and the $z=1-3$ galaxies from \citet{genzel2010}, there is a strong correlation (a Spearman's rank test gives a correlation coefficient of $\rho=0.72$ with a two-sided significance equal to 0.0). Our sample lies in the range expected. This suggests that we are not observing any differences between our starburst and main sequence galaxies in Figures \ref{fig:lir_lco} and \ref{fig:SFE} simply because we are not probing a large enough range of $L_{\rm CO}^\prime$ and $L_{\rm IR}$.

\subsection{Comparing Different Tracers of Star Forming Regions}
\label{sec:PAH}
IR, CO, and PAH luminosities are all commonly used
as tracers of star formation in dusty galaxies. PAH emission arises from PDRs
surrounding young stars and has been demonstrated locally to be largely cospatial with the molecular clouds traced by CO emission \citep{bendo2010}. 
If star formation is continuously fueled for 
$\lesssim1$\,Gyr, these tracers should all correlate.

For most star forming galaxies, the ratio of $L_{\rm PAH}/L_{\rm IR}$ is fairly constant, but there is an observed deficit of PAH emission relative to $L_{\rm IR}$ in 
local ULIRGs, possibly due to an increase in the hardness of the radiation 
field caused either by a major merger/starburst or an AGN \citep{tran2001,desai2007}. This same deficit does not hold for similarly luminous galaxies at high redshift, however, where the majority of ULIRGs are observed to have 
strong PAH emission \citep{pope2008,menendez2009,kirkpatrick2012}.

\citet{pope2013} explored the evolution of $L_{\rm 6.2}/L_{\rm IR}$ with redshift for a sample of ULIRGs from $z\sim1-4$ as well as a sample of local ULIRGs.
Specifically, the authors compare $L_{\rm 6.2}/L_{\rm IR}$ with $L_{\rm IR}$ and find that 
the deficit in $L_{\rm 6.2}$ relative to $L_{\rm IR}$ occurs at a higher $L_{\rm IR}$ for high redshift galaxies than is seen in the local Universe. 
Galaxies from $z\sim1-3$ typically have higher gas fractions than local counterparts \citep{daddi2010a,tacconi2010}. This increase in molecular gas could be linked to the
relative increase in PAH emission, since both are largely cospatial. Indeed, when \citet{pope2013} compare $L_{\rm 6.2}/L_{\rm IR}$ with
$L_{\rm IR}/L^\prime_{\rm CO}$, they find a consistent relationship for both the local and high redshift ULIRGs.

\begin{figure*}[ht!]
\plotone{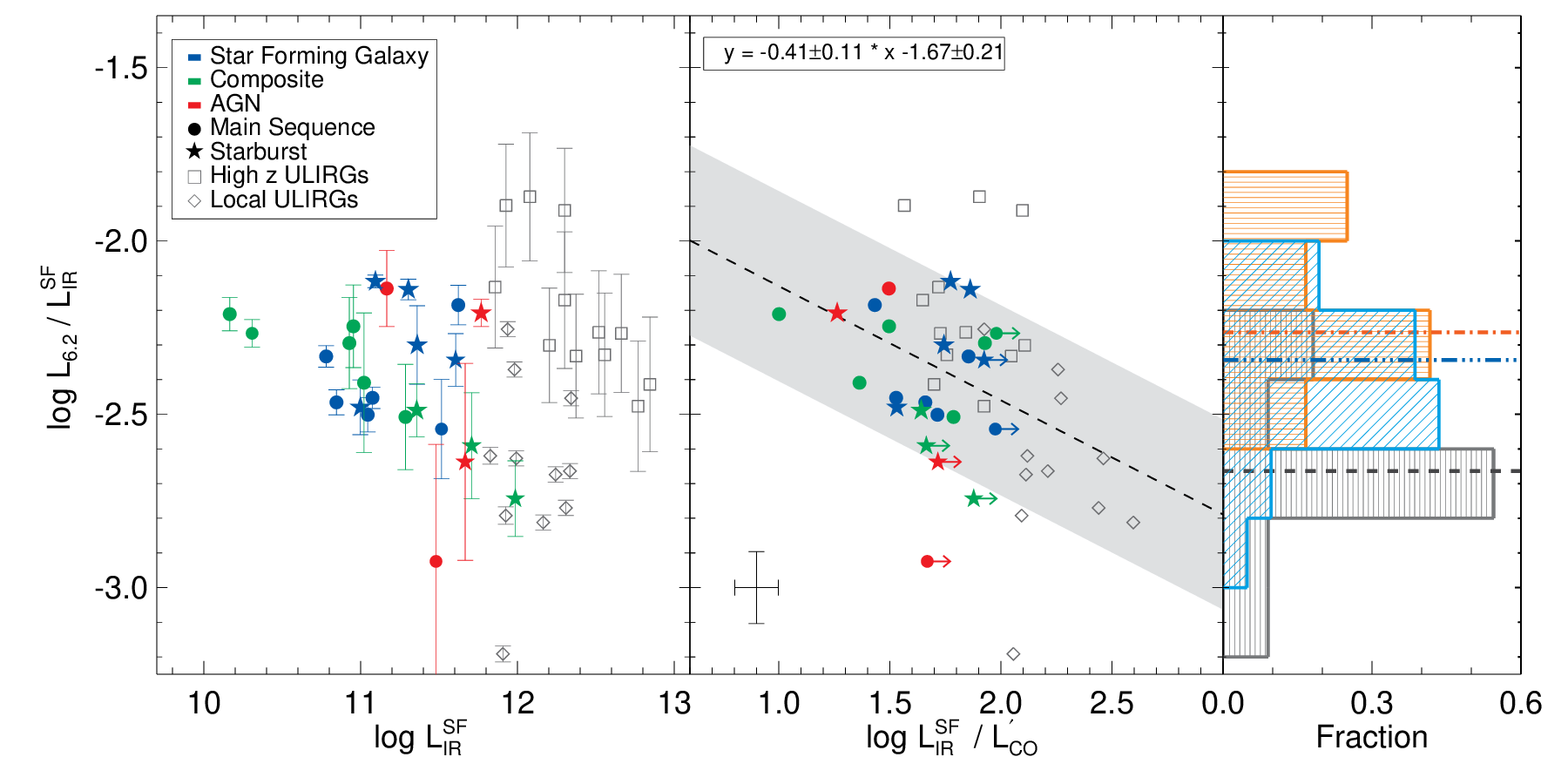}
\caption{We plot $L_{\rm 6.2}/L_{\rm IR}$ vs. $L_{\rm IR}^{\rm SF}$ in the {\it left panel} and vs. $L_{\rm IR}^{\rm SF}/L^\prime_{\rm CO}$ in the {\it middle panel}. We color the points
according to the power source of the mid-IR luminosity, and we use different symbols to designate starbursts and main sequence galaxies based on sSFR. We also plot as the grey points the local and high redshift ULIRGs from \citet{pope2013}. The 5MUSES galaxies combined with the local ULIRGs comprise a low redshift sample for comparison with the high redshift ULIRGs. The 5MUSES galaxies have similar $L_{\rm 6.2}/L_{\rm IR}^{\rm SF}$ ratios as the high redshift ULIRGs, while the local ULIRGs have a deficit, likely related to their more compact emission. When we normalize the dust emission by the molecular gas emission (middle panel), all galaxies lie in a similar region of parameter space, illustrating the consistent link between PAH emission and molecular gas over a range of $L_{\rm IR}$ and redshifts. For clarity, we omit individual error bars and plot the typical uncertainties for all galaxies in the lower left corner. The dashed line shows the best fit relation for all galaxies (listed in the upper left corner), and the grey shaded region marks the standard deviation about this line. We show the distributions in the histogram on the right. The grey histogram is the distribution of $L_{\rm 6.2}/L_{\rm IR}^{\rm SF}$ for the local ULIRGs, with the median overplotted as the dashed line; the orange histogram and line show the distribution and median for the high redshift ULIRGs, and the cyan histogram and line is the distribution and median for the 5MUSES sample. \label{fig:pah_co}}
\end{figure*}

We now build on the analysis presented in \citet{pope2013} by extending the parameter space explored to the lower luminosity 5MUSES sample. The 5MUSES galaxies combined with the local ULIRGs comprise a low redshift sample for comparison with the high redshift ULIRGs. We plot $L_{\rm 6.2}/L_{\rm IR}^{\rm SF}$ vs. $L_{\rm IR}^{\rm SF}$ for our sample in the left panel of Figure \ref{fig:pah_co}. We also include the high redshift and local ULIRGs from \citet{pope2013} where we have calculated $L_{\rm IR}^{\rm SF}$ for all ULIRGs by scaling the mid-IR AGN strength, determined by decomposing the mid-IR spectra. There is a decreasing trend between the 5MUSES sample and the local ULIRGs, but the high redshift ULIRGs are shifted from this relation. In the right panel, we plot $L_{\rm 6.2}/L_{\rm IR}^{\rm SF}$ vs. $L_{\rm IR}^{\rm SF}/L^\prime_{\rm CO}$. $L_{\rm CO}^\prime$ is calculated using the estimated CO(1-0) luminosity for all galaxies \citep[see][for conversion details]{pope2013}. 
In this panel, most galaxies follow the same decreasing trend, with a few obvious outliers. We overplot the best fit relation
for the 5MUSES sample and the local and high redshift ULIRGs. The shaded region indicates one standard deviation above and below the fit. There is a decreasing correlation between
$L_{\rm 6.2}/L_{\rm IR}^{\rm SF}$ and $L_{\rm IR}^{\rm SF}/L^\prime_{\rm CO}$ for most galaxies.

Figure \ref{fig:pah_co} suggests that the relative amount of emission from small dust grains is related to $L_{\rm IR}^{\rm SF}/L^\prime_{\rm CO}$ for dusty galaxies out to $z\sim2$. That is, weaker PAH emission is 
associated with a higher star formation efficiency and faster gas depletion timescales. The decrease in $L_{6.2}$ with increasing $L_{\rm IR}$ could indicate 
that PAH emission in general is suppressed for more luminous galaxies. We do not find significantly lower $L_{\rm 6.2}/L_{\rm IR}^{\rm SF}$ ratios for our AGN or composite galaxies as compared to our star 
forming galaxies, indicating that in our sample, the growing AGN is not affecting this ratio. 
 
As discussed in \citet{pope2013}, the PAH deficit could be similar to the observed deficit in [CII] emission at high 
$L_{\rm IR}$ in local galaxies \citep[e.g.,][]{kaufman1999,stacey2010,carpio2011}.
\citet{santos2013} probe the [CII] deficit in local LIRGs and find that galaxies with
compact mid-IR emission have a [CII] deficit, regardless of the mid-IR power source. For our sample, follow-up observations are required to trace the compactness of the galaxies. If our galaxies have extended dust emission, this would explain the similar $L_{6.2}/L_{\rm IR}^{\rm SF}$ ratios for the 5MUSES galaxies and the high redshift ULIRGs, since high redshift ULIRGs are known to have extended dust emission \citep[e.g.,][]{chapman2004}. Based on the relative strengths of the dust emission and CO(1-0) emission, the 5MUSES sources, primarily LIRGs, seem to be more accurate counterparts for the high redshift ULIRGs than the local ULIRGs, evidencing the evolution of ISM properties with redshift. A morphological comparison of these sources could provide more insight into structure and compactness of the dust and gas emission.

\subsection{Gas Fractions}
\label{sec:gas}
\begin{figure}
\plotone{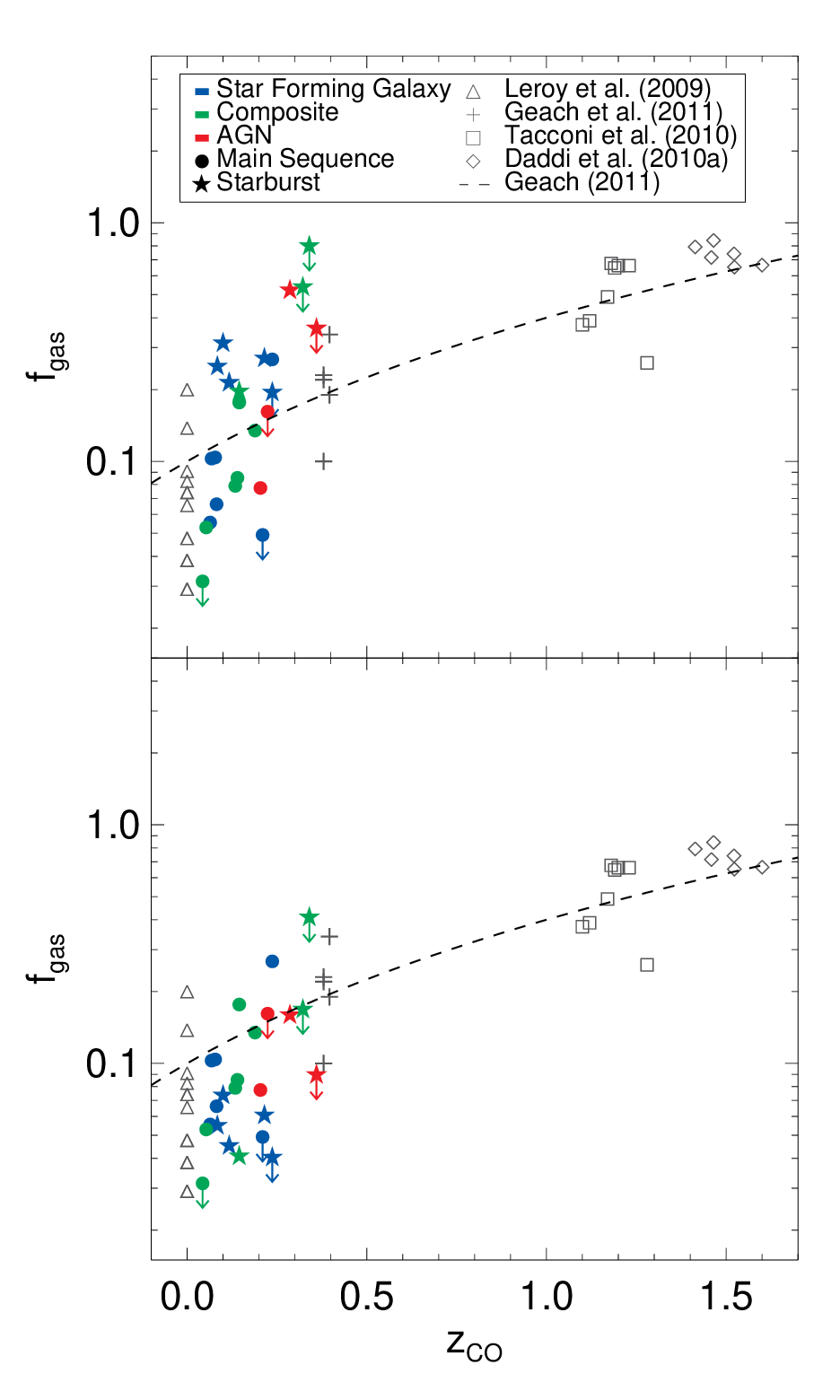}
\caption{Gas fractions vs. $z_{\rm CO}$ for our sample. {\it Top panel:} $f_{\rm gas}$ is calculated using $\alpha_{\rm CO}=4.6$ for all galaxies. {\it Bottom panel: } $f_{\rm gas}$ is calculated using $\alpha_{\rm CO} = 0.8$ for starburst galaxies (filled stars). We also overplot gas fractions from \citet{leroy2009}, \citet{daddi2010}, \citet{tacconi2010}, and \citet{geach2011}. The best fit line is $f_{\rm gas} = 0.1(1+z)^2$ \citep{geach2011}. Our galaxies lie in the region expected from the best fit line, although there is an offset between the starbursts (stars) and main sequence sources (circles) when $\alpha_{\rm CO}=4.6$ is used, indicating that a lower $\alpha_{\rm CO}$ conversion might be more appropriate for the starbursts. \label{fig:gas}}
\end{figure}

The gas fraction is expressed as $f_{\rm gas} = M_{\rm gas}/(M_{\rm gas} + M_\ast)$, where $M_{\rm gas} = \alpha_{\rm CO} L^\prime_{\rm CO}$. $\alpha_{\rm CO}=4.6$ is a commonly adopted value for normal star forming galaxies, while in starbursts, the conversion $\alpha_{\rm CO}=0.8$ has been measured \citep[and references therein]{bolatto2013}. 
We have two observational indicators of starburstiness: $L_{\rm IR}^{\rm SF}/L_{\rm CO}^\prime$ and sSFR.
To calculate $M_{\rm gas}$, we explore two scenarios.
First, we apply $\alpha_{\rm CO} = 4.6$ to our entire sample (top panel of Figure \ref{fig:gas}). None of our galaxies have $L_{\rm IR}^{\rm SF}/L_{\rm CO}^\prime$ indicative of a starburst (Figure \ref{fig:SFE}), so applying the same 
$\alpha_{\rm CO}$ to the entire sample is a reasonable assumption. 
Second, we use $\alpha_{\rm CO}=0.8$ to calculate the gas mass for those galaxies with an sSFR indicative of a starburst (bottom panel of Figure \ref{fig:gas}).
We plot the gas fractions as a function of redshift  We also plot gas fractions of normal star forming galaxies from the literature, and we have corrected the individual 
$\alpha_{\rm CO}$ values used to be 4.6. The average molecular gas fraction evolves with redshift, and we plot the measured relation, $f_{\rm gas}\propto(1+z)^2$, determined from a stellar mass limited sample with $\log M_\ast>10$ \citep{geach2011}, similar to the masses of our 5MUSES sample.

Our gas fractions (in both panels) lie in the range expected when comparing with the best fit line and the points from \citet{leroy2009} and \citet{geach2011}. Our CO(1-0) detection rate is high (17 out of 24 sources), producing a large range of measured gas fractions. In the top panel, our starburst galaxies are 
systematically higher than the main sequence galaxies, and the scatter about the $(1+z)^2$ line is larger, which suggests that the lower $\alpha_{\rm CO}=0.8$ conversion factor might be more appropriate for these sources if we expect similar gas fractions for main sequence and starburst galaxies. Theoretically, the conversion factor depends
on the geometry of the CO and H$_2$ distribution. When the CO emission is extended and not confined to molecular clouds, warm, and has a high surface density, as is the case in mergers, then the lower $\alpha_{\rm CO}$ value is appropriate 
\citep[][and references therein]{bolatto2013}. \citet{magnelli2012} measure $\alpha_{\rm CO}$ for a sample of high redshift main sequence and starburst galaxies and find an anti-correlation 
between $\alpha_{\rm CO}$ and sSFR, which they interpret as evidence that the mechanisms responsible for raising a galaxy off the main sequence must also affect the physical conditions within the star forming regions. We also find that the gas fractions reproduce a similar separation between sources as the main sequence criterion, linking $\alpha_{\rm CO}$ with the sSFR.
In contrast, neither $L_{\rm IR}^{\rm SF}/L_{\rm CO}^\prime$ nor $L_{\rm 6.2}/L_{\rm IR}^{\rm SF}$ shows any separation between the starburst and main sequence galaxies, likely due to the limited range being probed. These ratios, then, are relatively stable for galaxies of a limited mass and luminosity range.
\citet{narayanan2012} argue against a bimodal $\alpha_{\rm CO}$ conversion factor, and instead develop a fitting formula for the conversion factor that depends on the metallicity and CO line intensity. We currently lack metallicities for our sample, so we cannot directly apply the prescribed variable conversion factor. Given the continuous relationship between sSFR and SFR/$L_{\rm CO}^\prime$ evidenced in Figure \ref{fig:tdep}, a continuous, rather than bimodal, conversion factor based on galactic environment may be the most appropriate choice and would mean that all galaxies in our sample obey a similar relationship between the molecular gas and the stellar mass.

\section{Conclusions}
We present new LMT/RSR CO(1-0) detections for 24 intermediate redshift galaxies from the 5MUSES sample. We use {\it Spitzer} mid-IR spectra, available for all sources, to diagnose the presence and strength of an AGN. We removed the AGN contribution to $L_{\rm IR}^{\rm tot}$ and probe the star formation, gas, and dust emission using $L_{\rm IR}^{\rm SF}$, $L_{\rm CO}^\prime$, and $L_{6.2}$. We find
\begin{enumerate}
\item Removing the AGN contribution to $L_{\rm IR}^{\rm tot}$ results in a mean $L_{\rm IR}^{\rm SF}/L_{\rm CO}^\prime$ for our entire sample consistent with the mean $L_{\rm IR}/L_{\rm CO}^\prime$ derived for a large sample of star forming galaxies from $z\sim0-3$. For our four AGN sources, removing the AGN contribution produces a mean $L_{\rm IR}^{\rm SF}/L_{\rm CO}^\prime$ lower than the mean $L_{\rm IR}^{\rm SF}/L_{\rm CO}^\prime$ for our star forming galaxies or composites. We find that
$L_{\rm IR}^{\rm SF}/L_{\rm CO}^\prime$ is not strongly correlated with either the sSFR or the mid-IR power source over the range of luminosities probed. 
\item The average ratio of $L_{\rm 6.2}/L_{\rm IR}^{\rm SF}$ in our sample is similar to what is observed in high redshift ULIRGs rather than local ULIRGs. 
When we plot 
$L_{6.2}/L_{\rm IR}^{\rm SF}$ as a function of  $L_{\rm IR}^{\rm SF}/L_{\rm CO}^\prime$, we find all galaxies (local ULIRGs, our intermediate redshift 5MUSES sources, and high redshift ULIRGs) are consistent with the same declining relationship.
\item Our starbursts have gas fractions that are clearly offset from the main sequence galaxies if we apply a constant $\alpha_{\rm CO}$ to all galaxies which might indicate that the two populations require different $\alpha_{\rm CO}$ values. However, no dichotomy between starbursts and main sequence galaxies is evident when comparing other quantities that probe the ISM ($L_{\rm IR}^{\rm SF}$, $L_{\rm CO}^\prime$, or $L_{\rm 6.2}$).
\end{enumerate}

We plan to improve this study in the future by expanding the sample size with more CO(1-0) observations from the LMT, in order to further test how the gas and dust depends on $f_{\rm AGN,IR}$. We also plan to measure the 
morphologies of this sample to look for merging signatures and evidence of extended dust/gas emission.

\acknowledgements
We thank the anonymous referee for the helpful comments which have improved the clarity of this work.
This work would not have been possible without the long-term financial support from the Mexican Science and Technology Funding Agency, CONACYT (Consejo Nacional de Ciencia y Tecnolog\'{i}a) during the construction and early operational phase of the Large Millimeter Telescope Alfonso Serrano, as well as support from the the US National Science Foundation via the University Radio Observatory program, the Instituto Nacional de Astrof\'{i}sica, \'{O}ptica y Electr\'{o}nica (INAOE) and the University of Massachusetts (UMASS). 
The UMass LMT group acknowledges support from NSF URO and ATI grants (AST-0096854 and AST-0704966) for the LMT project and the construction of the RSR. 
AK would like to acknowledge support from a William Bannick Student Travel Grant. We are grateful to all the LMT observers from Mexico and UMass who took data for this project.
This work is based in part on observations made
with the {\it Spitzer Space Telescope}, which is operated by the Jet
Propulsion Laboratory, California Institute of Technology under
a contract with NASA, and the {\it Herschel Space Observatory}, a
European Space Agency Cornerstone Mission with significant
participation by NASA. This research has made use of the NASA/
IPAC Extragalactic Database (NED) which is operated by the
Jet Propulsion Laboratory, California Institute of Technology,
under contract with the National Aeronautics and Space Administration.

{}

\end{document}